\documentclass[sigconf]{acmart}

\usepackage{booktabs} 

\copyrightyear{2018}
\acmYear{2018}
\setcopyright{acmlicensed}
\acmConference[MSR '18]{15th International Conference on Mining Software Repositories }{May 28--29, 2018}{Gothenburg, Sweden}
\acmBooktitle{MSR '18: 15th International Conference on Mining Software Repositories , May 28--29, 2018, Gothenburg, Sweden}
\acmPrice{15.00}
\acmDOI{10.1145/3196398.3196430}
\acmISBN{978-1-4503-5716-6/18/05}

\usepackage[export]{adjustbox} 
\usepackage{amssymb,amsmath}

\usepackage{listings}

\definecolor{light-gray}{gray}{0.97}
\definecolor{gray}{rgb}{0.4,0.4,0.4}
\definecolor{darkblue}{rgb}{0.0,0.0,0.6}
\definecolor{cyan}{rgb}{0.0,0.6,0.6}

\lstnewenvironment{code} {\lstset{
}}{}

\lstnewenvironment{java} {\lstset{
  language=Java
}}{}

\lstnewenvironment{regex} {\lstset{
  alsoletter=\\()\{\}.,
  morekeywords={String, long, if, return, format,B,int,Math,log}
}}{}

\lstdefinelanguage{XML}
{
  morestring=[b]",
  morestring=[s]{>}{<},
  morecomment=[s]{<?}{?>},
  stringstyle=\color{black},
  identifierstyle=\color{darkblue},
  keywordstyle=\color{cyan},
  morekeywords={xmlns,version,type,xmlns:data,xmlns:xlink,height,width,preserveAspectRatio,viewBox,style,data,id,xlink,href,authors,date,time,annotation,artifact,path,x,y,role,rel,inline,iconpos,transition,icon,theme,class,src}
}

\lstnewenvironment{sql} {\lstset{
	morekeywords={SELECT, FROM, WHERE, ORDER, BY, ASC, LIKE, INNER, JOIN, ON} 
}}{}

\usepackage{tabularx}

\usepackage{enumitem}

\usepackage{calc}

\usepackage{multirow}

\usepackage{bm} 

\usepackage{amsthm}
\newtheoremstyle{definitiontight}
  {.5\baselineskip\@plus.2\baselineskip\@minus.2\baselineskip}
  {.6\baselineskip\@plus.2\baselineskip\@minus.2\baselineskip}
  {}
  {}
  {\bfseries}
  {.}
  {.5em}
  {}
\theoremstyle{definitiontight}
\newtheorem{definition}{Definition}[section]

\usepackage{amsmath}
\usepackage{amssymb} 
\usepackage{eucal} 

\usepackage{siunitx}
\usepackage{algpseudocode}
\usepackage{algorithm}

\usepackage[export]{adjustbox}

\definecolor{VeryLightGray}{rgb}{0.92,0.92,0.92}
\usepackage{colortbl}
\newcommand{\graybg}{\cellcolor{VeryLightGray}}

\begin{document}
\title{SOTorrent: Reconstructing and Analyzing the Evolution of Stack~Overflow Posts}

\author{Sebastian Baltes}
\orcid{0000-0002-2442-7522}
\email{research@sbaltes.com} 
\author{Lorik Dumani}
\email{dumani@uni-trier.de}
\affiliation{%
  \institution{University of Trier, Germany}
}

\author{Christoph Treude}
\email{christoph.treude@adelaide.edu.au}
\affiliation{%
  \institution{University of Adelaide, Australia}
}

\author{Stephan Diehl}
\orcid{0000-0002-4287-7447}
\email{diehl@uni-trier.de}
\affiliation{%
  \institution{University of Trier, Germany}
}


\begin{abstract}
Stack Overflow (SO) is the most popular question-and-answer website for software developers, providing a large amount of code snippets and free-form text on a wide variety of topics.
Like other software artifacts, questions and answers on SO evolve over time, for example when bugs in code snippets are fixed, code is updated to work with a more recent library version, or text surrounding a code snippet is edited for clarity.
To be able to analyze how content on SO evolves, we built \emph{SOTorrent}, an open dataset based on the official SO data dump.
\emph{SOTorrent} provides access to the version history of SO content at the level of whole posts and individual text or code blocks.
It connects SO posts to other platforms by aggregating URLs from text blocks and by collecting references from GitHub files to SO posts.
In this paper, we describe how we built \emph{SOTorrent}, and in particular how we evaluated 134 different string similarity metrics regarding their applicability for reconstructing the version history of text and code blocks.
Based on a first analysis using the dataset, we present insights into the evolution of SO posts, e.g., that post edits are usually small, happen soon after the initial creation of the post, and that code is rarely changed without also updating the surrounding text.
Further, our analysis revealed a close relationship between post edits and comments.
Our vision is that researchers will use \emph{SOTorrent} to investigate and understand the evolution of SO posts and their relation to other platforms such as GitHub.
\end{abstract}

%
%



\maketitle

\section{Introduction}



Stack Overflow (SO) is the most popular question-and-answer website for software developers.
As of December 2017, its public data dump~\citep{StackExchangeInc2017b} lists over 38 million posts and over 8 million registered users.
Many answers contain code snippets together with explanations~\citep{YangHussainOthers2016}.
Similar to other software artifacts such as source code files and documentation~\cite{Lehman1980, ChapinHaleOthers2001, MensDemeyer2008, GodfreyGerman2008}, text and code snippets on SO evolve over time, e.g., when the SO community fixes bugs in code snippets, clarifies questions and answers, and updates documentation to match new API versions.
Since the inception of SO in 2008, a total of 13.9 million SO posts have been edited after their creation---19,708 of them more than ten times. 
While many SO posts contain code, the evolution of code snippets on SO differs from the evolution of entire software projects: Most snippets are relatively short (on average 12 lines, see Section~\ref{sec:analysis-evolution}) and many of them cannot compile without modification~\cite{YangHussainOthers2016}.
In addition, SO does not provide a version control or bug tracking system for post content, forcing users to rely on the commenting function or additional answers to voice concerns about a post.

Recent studies have shown that developers use SO snippets in their software projects, regardless of maintainability, security, and licensing implications~\cite{BaltesKieferOthers2017, AnMloukiOthers2017, YangMartinsOthers2017,  GharehyazieRayOthers2017, AbdalkareemShihabOthers2017, XiaBaoOthers2017, FischerBottingerOthers2017, AcarBackesOthers2016}.
Assuming that developers copy and paste snippets from SO without trying to thoroughly understand them, maintenance issues arise.
For instance, it may later be more difficult for developers to refactor or debug code that they did not write themselves. Moreover, if no link to the SO post is added to the copied code, it is not possible to check the SO thread for a corrected or improved solution in case problems occur. 

The SO data dump keeps track of different versions of entire posts, but does not contain information about differences between versions at a more fine-grained level.
In particular, it is not trivial to extract different versions of the same code snippet from the history of a post to analyze its evolution.
To address these challenges, we present the open dataset \emph{SOTorrent}, which enables researchers to analyze the version history of SO posts at the level of whole posts and individual post blocks, and their relation to corresponding source code in GitHub repositories.
We also use this dataset to answer three research questions about the evolution of post content on SO, which are, to the best of our knowledge, not sufficiently answered yet: 
\emph{How do Stack Overflow posts evolve?} (RQ1), \emph{Which posts get edited?} (RQ2), and \emph{What is the temporal relationship between edits and comments?} (RQ3).

While answering the first two questions will further our understanding of the phenomenon of SO post evolution, the third question aims at finding a connection between post edits and other events on the SO platform.
We found that SO posts grow over time in terms of their number of text and code blocks, but the size of the individual blocks is relatively stable.
Many edits ($44.1\%$) just modify a single line of text or code, but only in $6.1\%$ of the cases are code blocks changed without also changing text content; post edits usually happen shortly after the creation of the post.
Our research suggests that comments and post edits are closely related: Some comments might trigger edits, others might be made in response to the edits.
The contribution of this work consists of the publicly available dataset \emph{SOTorrent}, the algorithms and techniques used for its construction, and an initial analysis of SO post evolution.

\section{The SOTorrent dataset}

\begin{figure}
\centering
\includegraphics[width=1\columnwidth,  trim=0.0in 0.0in 0.0in 0.0in]{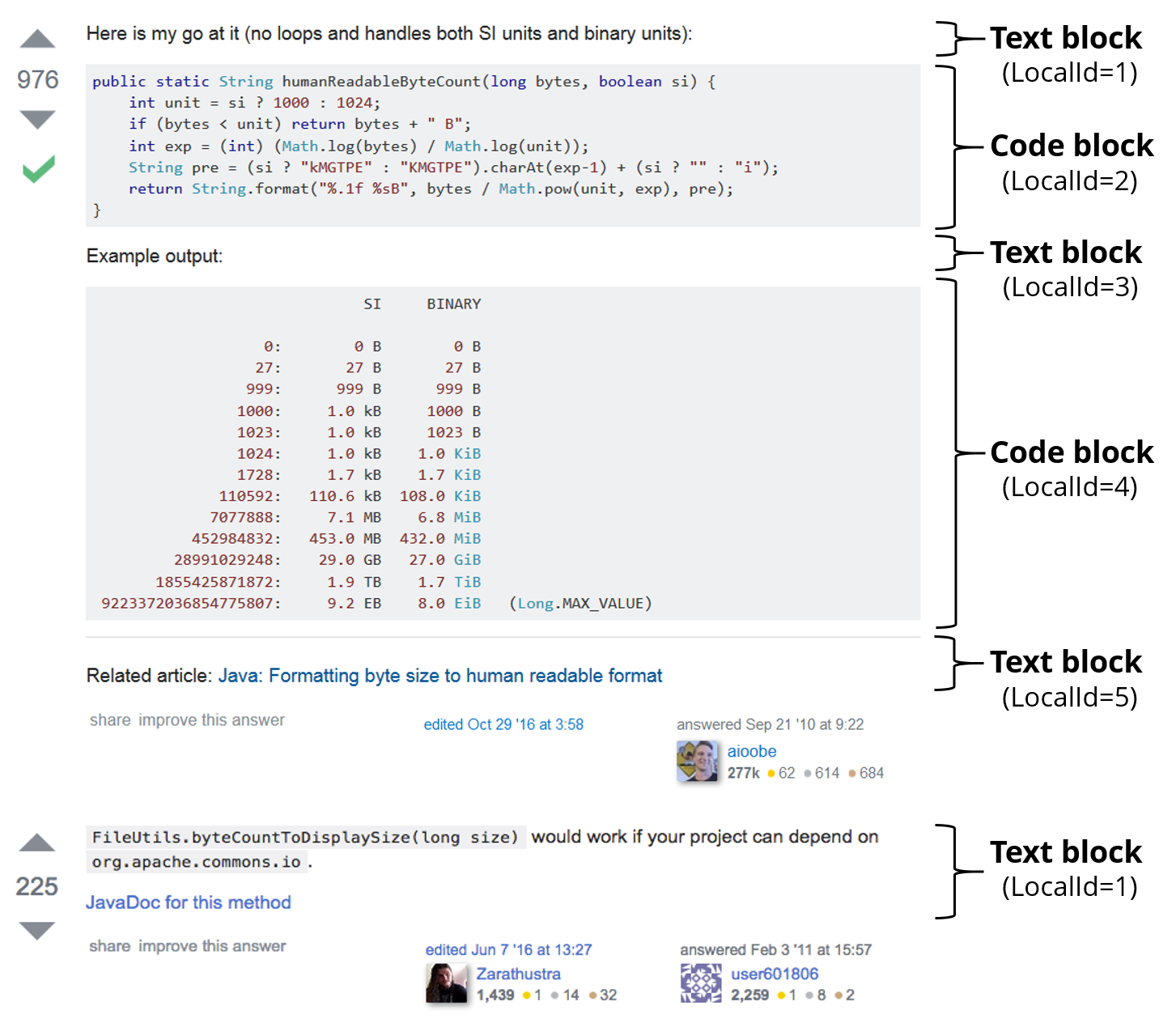} 
\caption{Exemplary Stack Overflow answers with code blocks (top, \href{https://stackoverflow.com/a/3758880}{3758880}) and with inline code (bottom, \href{https://stackoverflow.com/a/4888400}{4888400}). The \texttt{LocalId} represents the position in the post.}
\label{fig:so-postblocks-example}
\end{figure}

To answer our research questions, and to support other researchers in answering similar questions, we build \emph{SOTorrent}, an open dataset based on data from the official SO data dump~\cite{StackExchangeInc2017b} and the Google BigQuery GitHub (GH) dataset~\cite{GoogleCloudPlatform2018}.
\emph{SOTorrent} provides access to the version history of SO content at the level of whole posts and individual post blocks.
A post block can either be a text or a code block, depending on how the author formatted the content (see Figure~\ref{fig:so-postblocks-example} for an example).
Beside providing access to the version history, the dataset links SO posts to external resources in two ways: (1) by extracting linked URLs from text blocks of SO posts and (2) by providing a table with links to SO posts found in the source code of GitHub projects.
This table can be used to connect \emph{SOTorrent} and GH datasets such as \emph{GHTorrent}~\cite{Gousios2013}. 
Our dataset is available on Zenodo as a database dump~\cite{BaltesDumani2018} together with instructions on how to import the dataset. 
We also published the source code of the software that we used to build~\cite{BaltesDumani2018c, Baltes2018b} and analyze~\cite{Baltes2018d, Baltes2018e} \emph{SOTorrent}.

The current release \emph{2018-02-16} of the dataset contains the version history of all 38,394,895 questions and answers in the official SO data dump.
It contains 60,235,289 post versions and 186,924,947 post block versions, ranging from the creation of the first post on July 31, 2008 until the last edit on December 1, 2017. 
We extracted links to 11,019,477 distinct URLs from 19,453,365 different post block versions and further identified 5,816,307 links to SO posts in 430,521 public GH repositories. 
In the following sections, we provide information about \emph{SOTorrent}'s data storage and collection process, before we use the dataset to answer our research questions.

\section{Database Schema}
\label{sec:database-schema}

\begin{figure*}
\centering
\includegraphics[width=0.96\textwidth,  trim=2.7in 3.3in 2.7in 3.3in]{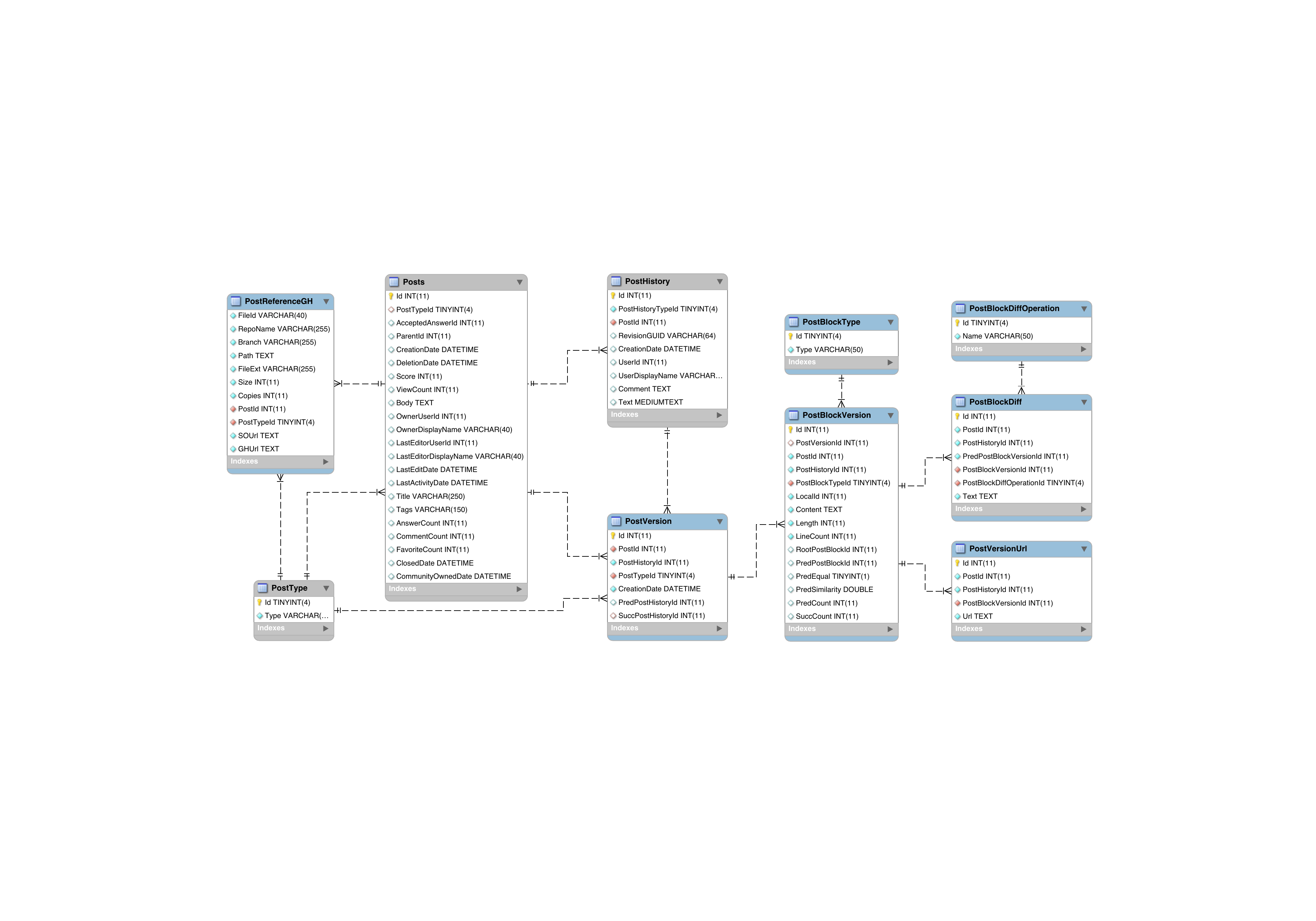} 
\caption{Database schema of \emph{SOTorrent}: The tables from the offical SO dump~\cite{StackExchangeCommunityWiki2018} are marked gray, the additional tables are marked blue. Not all tables from the official SO dump and not all foreign key constraints are shown.}
\label{fig:schema}
\end{figure*}

\emph{SOTorrent} contains all tables from the official Stack Overflow data dump, published December 1, 2017~\cite{StackExchangeInc2017b} (see database schema in Figure~\ref{fig:schema}).
However, that dump does only provide the version history at the level of whole posts as Markdown-formatted text.
To analyze how individual text or code blocks evolve, we needed to extract individual blocks from that content.
This extraction also enabled us to collect links to external sources from the identified text blocks.
In the SO dump, one version of a post corresponds to one row in the table \texttt{PostHistory}.
However, that table does not only document changes to the content of a post, but also changes to metadata such as tags or title.
Since our goal was to analyze the evolution of SO posts at the level of whole posts and individual post blocks, we had to filter and process the available data.
First, we selected edits that changed the content of a SO post, identified by their \texttt{PostHistoryTypeId}~\cite{StackExchangeCommunityWiki2018} ($2$: \textit{Initial Body}, $5$: \textit{Edit Body}, $8$: \textit{Rollback Body}).
We linked each filtered version to its predecessor and successor and stored it in table \texttt{PostVersion}.

The content of a post version is available as Markdown-formatted text.
We split the content of each version into text and code blocks (see Section~\ref{sec:postblock-extraction}) and extracted the URLs from all text blocks using a regular expression (table \texttt{PostVersionUrl}).
To reconstruct the version history of individual post blocks (table \texttt{PostBlockVersion}), we established a linear predecessor relationship between the post block versions using a string similarity metric that we selected after a thorough evaluation (see Section~\ref{sec:metrics-evaluation}).
For each post block version, we computed the line-based difference to its predecessor, which is available in table \texttt{PostBlockDiff}.

One row in table \texttt{PostReferenceGH} represents one link from a file in a public GH repository to a post on SO.
To extract those references, we utilized Google BigQuery, which allows to execute SQL queries on various public datasets, including a dataset with all files in the default branch of GitHub projects~\cite{GoogleCloudPlatform2018}.
To find references to SO, we applied the following regular expression to each line of each non-binary file in the dataset: 

\begin{regex}
(?i:https?://stackoverflow\.com/[^\s)\.\"]*)
\end{regex}

Because there are different ways of referring to questions and answers on SO, i.e. using full URLs or short URLs, we mapped all extracted URLs to their corresponding sharing link (ending with \verb+/q/<id>+ for questions and \verb+/a/<id>+ for answers) and stored that link together with information about the file and the repository in which the link was found in table \texttt{PostReferenceGH}.
We ignored other links referring to, e.g., users or tags on SO.

\section{Post Block Extraction}
\label{sec:postblock-extraction}

Our goal was to analyze the evolution of individual text and code blocks, for example to trace changes to particular code snippets or to identify bug fixes for code on SO.
Moreover, the differentiation between the two post block types allowed us to extract links to external resources only from text blocks, not from code blocks.
The latter may, for example, contain XML namespace links or links to stylesheet files, which we do not consider to be external sources of the post.
The first step towards reconstructing the version history of individual post blocks is their extraction from the Markdown-formatted text that SO uses for the content of posts.
In our notion, a code block is not a short inline code fragment embedded into a text block (see Figure~\ref{fig:so-postblocks-example} for an example), but a continuous code snippet.
We consider inline-code to be part of the surrounding text block.
According to SO's Markdown specification~\cite{StackExchangeInc2018}, code blocks are indented by four spaces and inline code is framed by backtick characters.
However, as we found during our research, users are free to use other Markdown specifications or HTML tags, which are not officially supported, but correctly parsed and displayed on the SO website.
We iteratively tested and refined our post block extraction approach using a random sample of over 100,000 SO posts ($s_\textit{large}$).
We ran the extraction, randomly checked the extracted posts blocks, and added a new test case if the result differed from the rendering on the SO website (class \texttt{PostVersionHistoryTest}~\cite{BaltesDumani2018c}).
We then updated the extraction such that all test cases passed and re-ran the extraction on the test data.
The final version of our post block extraction method was able to detect various notations that SO authors used to mark code blocks, including SO Markdown (indented by 4 spaces), code fencing Markdown (enclosed by three backticks), SO stack snippets (enclosed by \texttt{<!--begin/end snippet-->}), stack snippet language tags (prepended by \texttt{<!--language:...-->}), HTML code tags (enclosed by \texttt{<pre><code>}), and HTML script tags (enclosed by \texttt{<script>}).

\section{Post Block Matching}

After successfully extracting the post blocks from a post version, we had to map the extracted post blocks to their predecessors in the previous post version to reconstruct their version history.
Since this mapping had to work for text and code content, the latter in various programming languages, we decided to utilize syntax-based similarity metrics.
We implemented 134 different string similarity metrics (see Section~\ref{sec:similarity-metrics}), which we evaluated regarding their correctness and performance using the manually validated version history of 600 SO posts (see Sections~\ref{sec:ground-truth} and \ref{sec:metrics-evaluation}).
In case of multiple matches, we had to choose between different predecessor candidates.
Thus, we developed a matching strategy that considers the location and context of a post block (see Section~\ref{sec:matching-strategy}).

\subsection{Similarity Metrics}
\label{sec:similarity-metrics}

\begin{table*}
\caption{Overview of all evaluated similarity metrics ($n=134$).}
\vspace{-0.4\baselineskip}
\small
\begin{tabular}{llll}
\hline\noalign{\smallskip}
\textbf{Type} & \multicolumn{2}{l}{\textbf{Metric}} & \textbf{Variants} \\
\noalign{\smallskip}\hline\noalign{\smallskip}
\multirow{2}{*}{edit} & levenshtein & damerauLevenshtein & \multirow{2}{*}{\shortstack[l]{with/without normalization}}\\
& longestCommonSubsequence (LCS) & optimalAlignment (OA) & \\
\noalign{\smallskip}\hline\noalign{\smallskip}
\multirow{2}{*}{set} & nGram\{Jaccard|Dice|Overlap\} & nShingle\{Jaccard|Dice|Overlap\} & \multirow{2}{*}{\shortstack[l]{$n\text{Gram} : n \in \{2,3,4,5\}$, $n\text{Shingle} : n \in \{2,3\}$\\ with/without normalization, padding (nGram)}}\\
& token\{Jaccard|Dice|Overlap\} & \\
\noalign{\smallskip}\hline\noalign{\smallskip}
\multirow{3}{*}{profile} & cosineNGram\{Bool|TF|NormalizedTF\} & manhattanNGram  & \multirow{3}{*}{\shortstack[l]{$n\text{Gram} : n \in \{2,3,4,5\}$, $n\text{Shingle} : n \in \{2,3\}$\\ with normalization (both) and without (cosine)}}\\
& cosineNShingle\{Bool|TF|NormalizedTF\} & manhattanNShingle & \\
& cosineToken\{Bool|TF|NormalizedTF\}  & manhattanToken & \\
\noalign{\smallskip}\hline\noalign{\smallskip}
\multirow{2}{*}{fingerprint} & \multicolumn{2}{l}{\multirow{2}{*}{winnowingNGram\{Jaccard|Dice|Overlap|LCS|OA\}}} & \multirow{2}{*}{\shortstack[l]{$n\text{Gram} : n \in  \{2,3,4,5\}$, \\ with/without normalization}}\\
&  & \\
\noalign{\smallskip}\hline\noalign{\smallskip}
\multirow{1}{*}{equal} & equal & tokenEqual & \multirow{1}{*}{\shortstack[l]{with/without normalization}}\\
\noalign{\smallskip}\hline
\end{tabular}
\label{tab:similarity-metrics}
\end{table*}

A similarity metric maps two input strings to a value in $[0, 1]$, where $0$ corresponds to  inequality and $1$ corresponds to equality.
We implemented five different types of similarity metrics: \emph{edit-based metrics} (e.g., Levenshtein), \emph{set-based metrics} (e.g., n-grams with Jaccard coefficient), \emph{profile-based metrics} (e.g, cosine similarity), \emph{fingerprint-based metrics} (Winnowing), and \emph{equality-based metrics}, which served as a baseline in the metrics evaluation (see Section~\ref{sec:metrics-evaluation}).
Our Java implementation of all metrics is available on GitHub~\cite{BaltesDumani2018d}.
Table~\ref{tab:similarity-metrics} shows all metrics that we implemented and evaluated.

The \emph{edit-based metrics} define the similarity of two strings based on the number of edit operations needed to transform one string into the other.
Optimal string alignment (OA) allows the two operations `insertion of one character' and `deletion of one character'.
The Levenshtein distance further allows `substitution of one character'.
The Damerau-Levenshtein distance is similar to Levenshtein, but additionally allows the operation `swap two neighboring characters'.
The longest common subsequence (LCS) of two strings is the longest sequence of characters (order irrelevant) that can be found in both strings.
It can be interpreted as a variant of Damerau-Levenshtein with the additional restriction that each character can only be modified once (e.g., swapping two characters and then replacing one of them is not possible).
To derive a similarity metric from the number of edit operations and the longest common subsequence, we used the following approaches:

\begin{definition}[Edit/LCS Similarity]
Let $S_1$, $S_2$ be two strings, $d$ be the edit distance and $LCS$ be the longest common subsequence between the two strings: $(S_1, S_2)  \to \mathbb{R}_{0}^{+}$. The edit- and LCS-based similarity functions $sim \colon (S_1, S_2) \to [0, 1]$ are then defined as
\[ sim_\text{edit}(S_1, S_2) = \frac{max(|S_1|, |S_2|) - d(S_1, S_2)}{max(|S_1|, |S_2|)} \]
\[ sim_\text{lcs}(S_1, S_2) = \frac{LCS(S_1, S_2)}{max(|S_1|, |S_2|)} \]
\end{definition}

The \emph{profile-based metrics} consider each distinct token, n-gram, or n-shingle in the two input strings as one dimension of a vector space.
Tokens can be extracted from a string by a tokenization with whitespaces as delimiter, $n$-grams split the string in sequences of $n$ consecutive characters, $n$-shingles split the string in sequences of $n$ consecutive words or tokens.
One input string is then characterized as one vector in the vector space.
In the simplest form (bool), the values of the dimensions can either be 1 (token, n-gram, or n-shingle present in the string) or 0 (not present).
Alternatively, one can consider the number of occurrences of each token, n-gram, or n-shingle as the value of the dimensions (term frequency).
We also considered the BM15 weighting scheme ($k=1.5$)~\cite{ManningRaghavanOthers2008}, which intends to lower the effect of very frequent terms skewing the comparison.
The similarity of the two strings is then defined as the cosine or Manhattan distance between the two vectors that have been derived from the strings using one of the three approaches described above.

For the \emph{set-based metrics}, we considered all distinct tokens, n-grams and n-shingles in the strings as elements of sets. We used three coefficients to compare the resulting sets:

\begin{definition}[Similarity Coefficients]
Let $S_1$, $S_2$ be sets of tokens, n-grams, or n-shingles.
\begin{align*}
Jaccard(S_1, S_2) &= \frac{|S_1 \cap S_2|}{|S_1 \cup S_2|} \hspace{2.2em} Dice(S_1, S_2) = \frac{2 \cdot |S_1 \cap S_2|}{|S_1| + |S_2|} \\
Overlap(S_1, S_2) &= \frac{|S_1 \cap S_2|}{min(|S_1|, |S_2|)} & \\
\end{align*}
\end{definition}

\begin{figure}
\centering
\includegraphics[width=1\columnwidth,  trim=0.0in 0.95in 0.0in 0.0in]{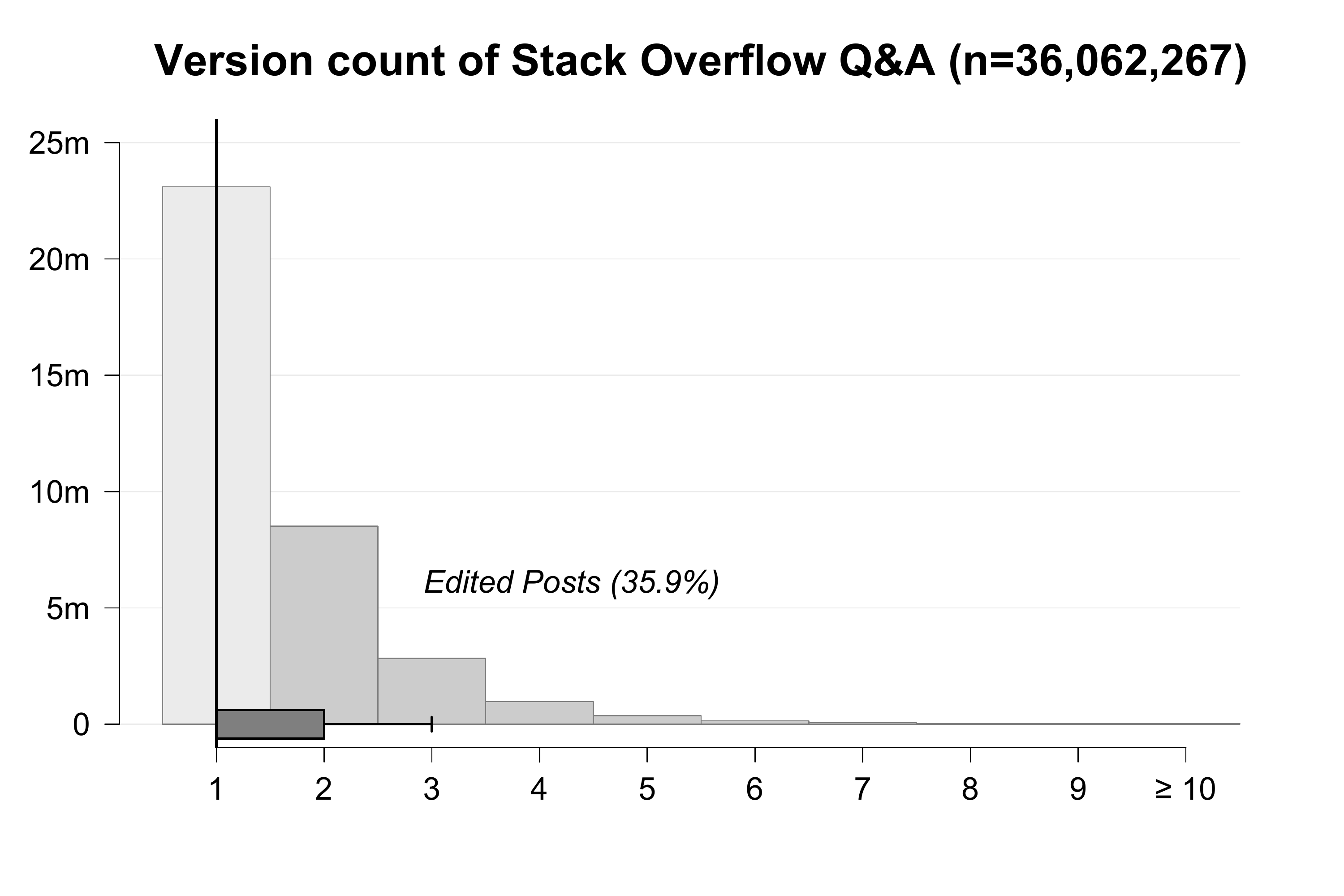} 
\caption{Histogram and boxplot showing the number of Stack Overflow questions and answers with a certain version count (PostHistoryTypeIds 2, 5, 8); based on the SO data dump 2017-06-12; vertical line is median.}
\label{fig:version-count}
\end{figure}

The \emph{fingerprint-based metrics} apply a hash function to substrings of the input strings and then use the computed hash values to determine the similarity.
The Winnowing algorithm is one approach to calculate and compare the fingerprints of two strings~\cite{SchleimerWilkersonOthers2003, DuricGasevic2013}.
Winnowing is often used for plagiarism detection, e.g., in the source code comparison software \emph{MOSS}~\cite{BurrowsTahaghoghiOthers2007, MartinsFonteOthers2014, LancasterCulwin2004}.
We implemented different variants of the algorithm described by Schleimer et al.~\cite{SchleimerWilkersonOthers2003}, e.g., using different n-grams sizes and different approaches to compare the fingerprints.

We implemented each metric in different variations.
In the variants with normalized input strings, we used different approaches for different metric types:
For the edit metrics, we unified the whitespace characters, i.e. reduced them to a single space, and converted all characters to lower case.
For the n-gram metrics, we converted all characters to lower case, removed all whitespace, and removed some special characters (\texttt{\{\};}).
For the shingle metrics, we again converted all characters to lower case, unified the whitespace characters, and removed all non-word characters (\texttt{[\^{}a-zA-Z\_0-9]}). 
We used common n-gram and shingle sizes~\cite{BurrowsTahaghoghiOthers2007} and also implemented an optional n-gram padding that emphasizes the beginning and the end of the input strings.
All these variations lead to a total number of 134 different similarity metrics.

\begin{figure*}
\centering
\includegraphics[width=0.95\textwidth,  trim=0.0in 0.2in 0.0in 0.0in]{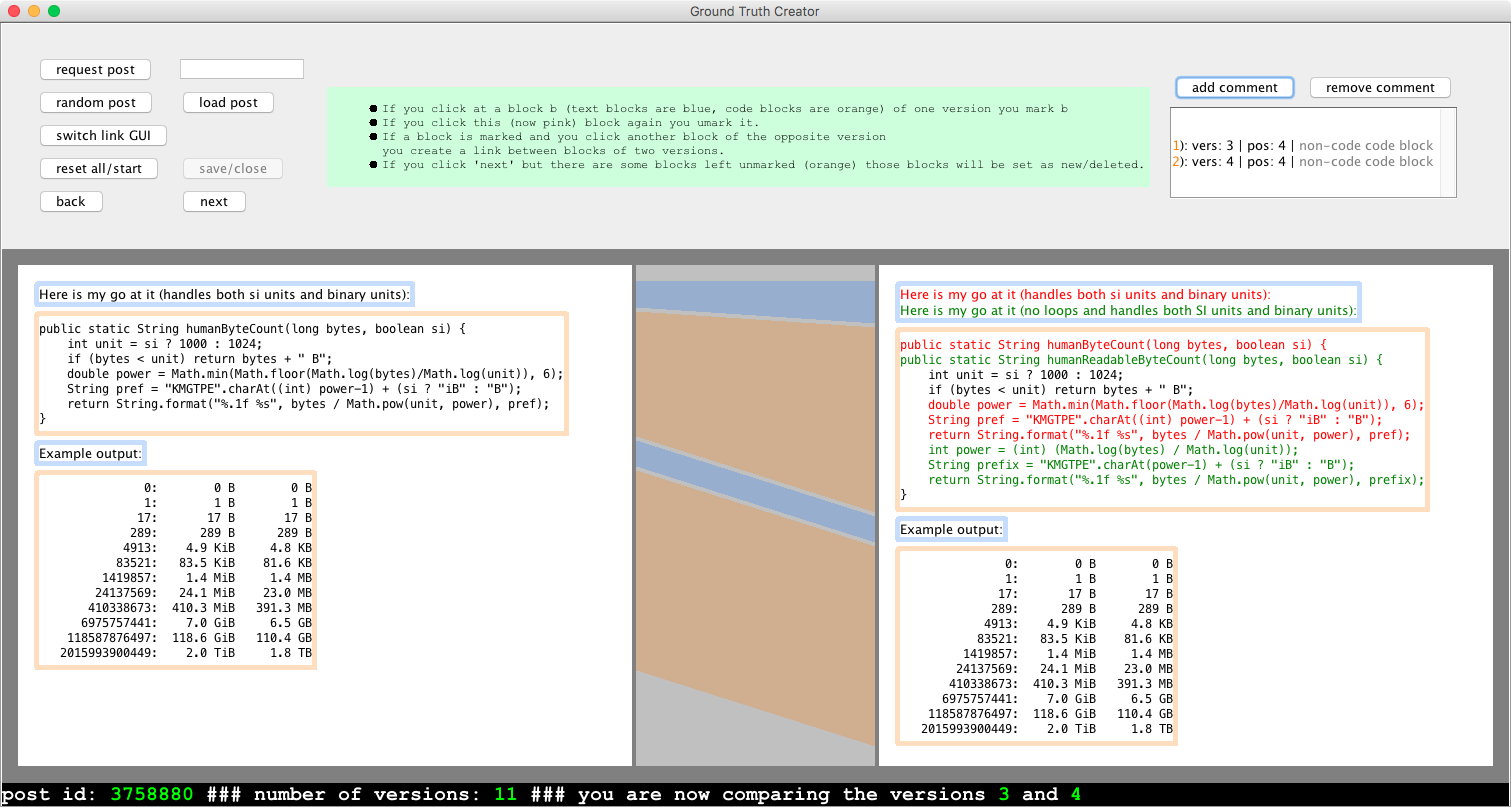} 
\caption{App developed to create ground truth for similarity metric evaluation.}
\label{fig:gt-app}
\end{figure*}

\begin{figure}
\centering
\includegraphics[width=1\columnwidth,  trim=0.0in 0.0in 0.0in 0.0in, frame]{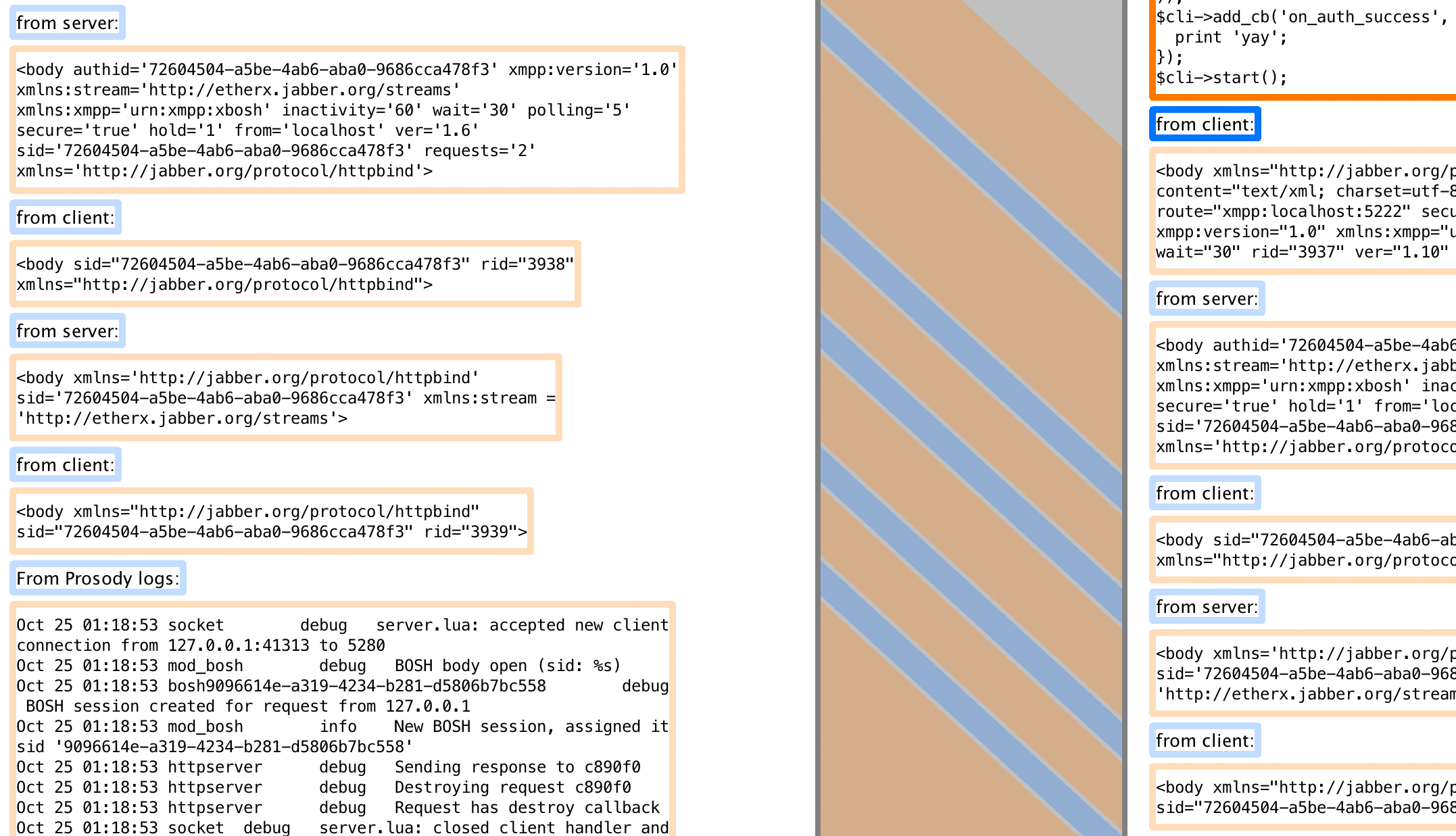} 
\caption{Post with multiple equal predecessors (\href{https://stackoverflow.com/posts/13064858/revisions}{13064858}).}
\label{fig:multiple-equal-predecessors}
\end{figure}

\subsection{Ground Truth}
\label{sec:ground-truth}

To evaluate the correctness of the post block mappings retrieved using different string similarity metrics, we created a set of 600 manually validated post version histories.
Figure~\ref{fig:gt-app} shows a screenshot of the tool we developed to create those manually validated histories (available on GitHub~\cite{DumaniBaltes2017}).
It visualizes a post version (right) and its predecessor (left).
Post blocks with equal content and type that are unique in the two versions are automatically connected.
For the other post blocks, the user has to choose a match by clicking on a post block of the same type in each version; the tool then visualizes the line-based difference between the connected blocks.
It is also possible to add comments for individual post blocks, e.g., in case the user is not confident in his or her mapping, or in case the post block extraction failed. 

We drew four different samples from the SO data dump released June 12, 2017.
The first sample with 200 posts ($s_\textit{rand}$) was randomly drawn from all SO questions and answers with at least two versions (otherwise no mapping is needed).
Since there are many posts with only two versions (see Figure~\ref{fig:version-count}), we decided to draw another sample of 200 posts from SO questions and answers with at least seven versions (99\% quantile).
We denote this sample $s_\textit{rand+}$.
As the initial focus of our research was on Java, we also drew a sample with 200 Java posts ($s_\textit{java}$) from all SO questions tagged with \texttt{<java>} or \texttt{<android>}, and the corresponding answers.
The last sample ($s_\textit{mult}$), which contains 100 posts with multiple possible predecessors, was not used to evaluate the metrics, but to evaluate our matching strategy (see Section~\ref{sec:matching-strategy}).
In this sample, we included posts which had at least two possible matches (two post blocks of the same type with identical content) in two adjacent versions.

The validated version histories of the samples were created by a graduate student, and then later discussed with two of the authors.
The student was introduced to the app and told to comment all post blocks where he is not sure about the mapping. 
Together, we looked at all post blocks with comments indicating an unclear mapping ($n=38$) and tried to find a mapping we all agreed on.
If that was not possible, we moved the post to a new sample $s_\textit{unclear}$, which we separately analyzed.
After discussing all 38 posts, $s_\textit{unclear}$ contained 17 posts (4 from $s_\textit{rand}$, 8 from $s_\textit{rand+}$, and 5 from $s_\textit{java}$).
All samples are available on Zenodo~\cite{BaltesDumaniOthers2017}.

\subsection{Matching Strategy}
\label{sec:matching-strategy}

Our goal was to establish a linear predecessor relationship for all post block versions, thus each post block version can only have one predecessor.
The reason for this decision was the fact that we rarely observed splits and merges in the post version histories we manually analyzed.
Moreover, even if multiple predecessors have equal or similar content, usually only one of them is the actual predecessor (see Figure~\ref{fig:multiple-equal-predecessors} for an example).
To correctly choose the predecessor from different candidates, we had to develop a matching strategy for post block versions, which we present in this section.
In the database, we not only store the matched predecessor, but also the number of possible predecessors and successors, to be later able to identify post version histories that could contain splits or merges. 
For our analysis (see Section~\ref{sec:analysis}), we consider \textit{post block lifespans}, i.e. chains of connected post block versions that are predecessors of each other.
Those lifespans can be easily retrieved from the database, because each post block version has a \texttt{RootPostBlockId}, which is the id of the first post block version in the chain.
As mentioned above, we utilized a dedicated sample $s_\textit{mult}$ to evaluate how well our matching strategy can handle posts with multiple possible connections.
In case of differences between the ground truth and the results of our approach, we wrote unit tests replicating the issue and then updated the strategy until all unit tests passed.
We further used the sample $s_\textit{large}$ to test the strategy's scalability.
To be able to describe our matching strategy, we define our notation for post versions, post block versions, and possible predecessors:

\begin{definition}[Post Version]
Let $p$ be a post with $n$ versions.
Then $\,p_{i}$ denotes one post version and $|p_i|$ denotes the number of post blocks in $p_i$ for $i \in \{1 \ldots n\}$.
\end{definition}

\begin{definition}[Post Block Version]
Let $p_i$ be one post version and $\tau \in \{\text{text},~\text{code}\}$ be a post block type.
Then $\,b_{(i, l)}^{\tau}$ denotes one post block of type $\tau$ with local id $\,l$ for $\,l \in \{1 \ldots |p_i|\}$.
The function $id^\tau \colon p_i \to \{ 1 \le l \le |p_i| \}$ maps a post version to the local ids of the post blocks of type $\tau$ in that version.
\end{definition}

\begin{definition}[Possible Predecessors]
Let $b_{(i-1, l)}^{\tau}$, $b_{(i, j)}^{\tau}$ be post blocks of the same type in subsequent post versions,
\[ equal(b_{(i-1, l)}^{\tau} , b_{(i, j)}^{\tau}) \to \{ \text{true}, \text{false} \} \]
be a function that tests if the post blocks' contents are equal, and
\[ sim^\tau(b_{(i-1, l)}^{\tau} , b_{(i, j)}^{\tau}) \to [0,1] \]
be the similarity of the two post blocks' contents according to the similarity metric $sim^\tau$.
Let $\vartheta^\tau \in [0,1]$ be a threshold for $sim^\tau$.
Then, we define the set of equal predecessors as
\begin{align*}
PredEqual(b_{(i, j)}^{\tau})=\{ \beta^{\tau}_{(i-1, k)}~|~&equal(\beta^{\tau}_{(i-1, k)}, b_{(i, j)}^{\tau}) = \text{true},\\
& k \in id^\tau(p_{i-1}),~j \in id^\tau(p_{i}) \}
\end{align*}
We define the maximum predecessor similarity as
\begin{align*}
maxSim^\tau = max(\{ sim^\tau(&\beta^{\tau}_{(i-1, k)} , b_{(i, j)}^{\tau})~|~sim^\tau \ge \vartheta^\tau, \\
& k \in id^\tau(p_{i-1}),~j \in id^\tau(p_{i}) \} )
\end{align*}
In case no predecessor with a similarity above the threshold exists, we define $maxSim^\tau(\emptyset) = 0$. 
We define the set of similar predecessors as
\begin{align*}
PredSim(b_{(i, j)}^{\tau})=\{ \beta^{\tau}_{(i-1, k)}~|~sim^\tau(&\beta^{\tau}_{(i-1, k)} , b_{(i, j)}^{\tau}) \ge maxSim^\tau, \\
& k \in id^\tau(p_{i-1}),~j \in id^\tau(p_{i}) \}
\end{align*}
Finally, we define the set of possible predecessors as
\[ Pred(b_{(i, j)}^{\tau}) =
\begin{cases}
PredEqual(b_{(i, j)}^{\tau}), &\mbox{if } PredEqual(b_{(i, j)}^{\tau}) \ne \emptyset , \\
PredSim(b_{(i, j)}^{\tau}), &\mbox{if } PredEqual(b_{(i, j)}^{\tau}) = \emptyset .
\end{cases} \]
The set of possible successors $Succ(b_{(i, j)}^{\tau})$ is defined analogously.
\end{definition}

As it can be seen in the above definition, we need two different similarity metrics ($sim^\textit{text}$ and $sim^\textit{code}$) and two different similarity thresholds ($\vartheta^\textit{text}$ and $\vartheta^\textit{code}$).
We only compute the similarity if the content of the post blocks is not equal, because we want to be able to distinguish equal post block versions from post block versions with similarity $1$ according to the metric.
Before we describe our matching strategy, we present two methods that we use in case of multiple possible predecessors.
Both methods iterate over all post blocks $b_{(i, j)}^{\tau}$ in a post version $p_{2 \le i \le n}$ that do not have a predecessor yet.
They follow different strategies for selecting a predecessor:

$setPredContext(p_i, BOTH)$ tries to select a predecessor using the post blocks before and after $b_{(i, j)}^{\tau}$, i.e. the blocks with local ids $j-1$ and $j+1$.
Please note that those blocks usually have a different post block type than $b_{(i, j)}^{\tau}$.
In case the predecessors of those neighboring blocks are already set and one post block $b_{(i-1, l)}^{\tau} \in Pred(b_{(i, j)}^{\tau})$ has the predecessors of those two post blocks as neighbors (local ids $l-1$ and $l+1$ in version $p_{i-1}$), the function sets $b_{(i-1, l)}^{\tau}$ as predecessor of $b_{(i, j)}^{\tau}$ and returns \textit{true}.
If no predecessor has been set, it returns \textit{false}.
In case of parameter \textit{ABOVE}, only the post block above (local id $j-1$) is taken into account; in case of parameter \textit{BELOW}, only the post block below (local id $j+1$) is taken into account.
Examples for posts that motivated this strategy are answer \href{https://stackoverflow.com/posts/32841902/revisions}{32841902} (mapping of version 2 to 1) and answer \href{https://stackoverflow.com/posts/37196630/revisions}{37196630} (mapping of version 2 to 1).

$setPredPosition(p_i)$ sets the post block $b_{(i-1, l)}^{\tau} \in Pred(b_{(i, j)}^{\tau})$ with $\Delta_\text{pos} = min(|l - j|)$, i.e. the post block with the local id closest to $j$, as predecessor of $b_{(i, j)}^{\tau}$.
If two possible predecessors have the same $\Delta_\text{pos}$, the method chooses the one with the smallest local id.
This approach is based on our observation that the order of post blocks rarely changes (see Section~\ref{sec:analysis-evolution}).
Examples for posts that motivated this strategy are question \href{https://stackoverflow.com/posts/18276636/revisions}{18276636} (mapping of version 2 to 1) and answer \href{https://stackoverflow.com/posts/2581754/revisions}{2581754} (mapping of version 3 to 2).

The complete matching strategy that selects (at most) one predecessor for each post block in a post version can be found as pseudo code in Algorithm~\ref{alg:matching-strategy}. 
The actual source code can be found in method \texttt{processVersionHistory} of class \texttt{PostVersionList} in the corresponding GitHub project~\cite{BaltesDumani2018c}.

\vspace{-0.5\baselineskip}
\begin{algorithm}
\caption{Matching Strategy}
\label{alg:matching-strategy}
\begin{algorithmic}
\ForAll{$p_{2 \le i \le n}$}
	\State \textit{// set predecessors where only one candidate exists}
	\ForAll{$b_{(i, 1 \le j \le |p_i|)}^{\tau}$}
		\If{$|Pred(b_{(i, j)}^{\tau})| = 1$}
			\State Let $pred$ be the equal or similar predecessor
			\If{$|Succ(pred)| = 1$}
				\State Set $pred$ as predecessor of $b_{(i, j)}^{\tau}$
				\State \textbf{continue}
			\EndIf
		\EndIf
	\EndFor
	\State \textit{// set predecessors using context}
	\State $predSet = \text{true}$
	\While{$predSet$}
		\State $predSet = setPredContext(p_i, BOTH)$
	\EndWhile
	\While{$predSet$}
		\State $predSet = setPredContext(p_i, BELOW)$
	\EndWhile
	\While{$predSet$}
		\State $predSet = setPredContext(p_i, ABOVE)$
	\EndWhile
	\State \textit{// set predecessors using position}
	\State $setPredPosition(p_i)$
\EndFor
\end{algorithmic}
\end{algorithm}

\subsection{Metrics Evaluation}
\label{sec:metrics-evaluation}

The matching strategy described above depends on the results of the similarity metrics $sim^\textit{text}$ and $sim^\textit{code}$ and the thresholds $\vartheta^\textit{text}$ and $\vartheta^\textit{code}$.
To select the best metrics for reconstructing the version history of post blocks, we evaluated all 134 metrics in different combinations with different thresholds using our ground truth samples $s_\textit{rand}$, $s_\textit{rand+}$, and $s_\textit{java}$.
Please note that the correctness of $sim^\textit{text}$ and $sim^\textit{code}$ cannot be evaluated independently, because the neighboring post blocks that $setPredContext$ takes into account usually have different types.
To assess the performance, we measured the runtime of the post history extraction for each configuration.
To assess the correctness of the extracted post block history, we regarded each metric configuration as a binary classifier that either assigns the predecessor of a post block version correctly or not (compared to the ground truth).
To calculate the number of true/false positives/negatives, we consider the set of \emph{predecessor connections}, i.e. all ($b_{(i-1, l)}^{\tau}$, $b_{(i, j)}^{\tau}$) from $\;p_{2 \le i \le n}$ that have been connected with a certain metric configuration.
We then compare those connections with the connections from the ground truth:

\begin{definition}[Metric Evaluation]
Let $\text{GT}^\tau$ be the set of predecessor connections of type $\tau$ in the ground truth, $\text{C}^\tau$ be the set of predecessor connections of type $\tau$ determined using a certain metric configuration,
and $\;n_\textit{pos}^\tau = \sum_{2 \le i \le n} |id^\tau(p_i)|\;$ be the number of possible predecessor connections of type $\tau$.
We define the number of true positives $\textit{tp}^\tau$, false positives $\textit{fp}^\tau$, true negatives $\textit{tn}^\tau$, and false negatives $\textit{fn}^\tau$ as:
\begin{align*}
\textit{tp}^\tau &= |\text{C} \cap \text{GT}| & \quad \textit{fp}^\tau &= |\text{C} \setminus \text{GT}|\\
\textit{tn}^\tau &= n_\textit{pos}^\tau - |\text{C} \cup \text{GT}| & \quad \textit{fn}^\tau &= |\text{GT} \setminus \text{C}|\\
\end{align*}
\end{definition}

After each comparison run, we ranked the configurations according to their Matthews correlation coefficient ($MCC$)~\cite{Matthews1975}, which takes $\textit{tp}^\tau$, $\textit{fp}^\tau$, $\textit{tn}^\tau$, and $\textit{fn}^\tau$ into account.
If two configurations had the same $MCC$ value, we ranked them according to their runtime.
$MCC$ is the preferred measure when evaluating binary classifiers~\cite{Chicco2017} and should be chosen over evaluation measures such as recall, precision, or F-measure~\cite{Powers2011}.
In our case, it correlates the connections from the ground truth and the connections set by a certain metric configuration.
The $MCC$ values are in range $[-1,1]$; a total disagreement is represented by $-1$, a perfect agreement by $1$.
The source code of the tool we used for the metrics evaluation is available on GitHub~\cite{BaltesDumani2018b}.

In the first comparison run, we configured $sim^\textit{text} = sim^\textit{code}$ and chose $\vartheta^\textit{\{text, code\}} \in \{0.0, 0.1, 0.2, \ldots, 1.0\}$.
This resulted in 1,474 different configurations.
The first run took about 24 hours on a regular desktop PC (Intel Core i7-7700, 64 GB RAM, 512 GB SSD). 

For the second run, we selected the metrics which, for a particular threshold, achieved a $MCC$ value in the 95\% quantile of all three samples either for text or for code blocks.
Some metrics cannot be applied to very short strings (e.g., if string length < n-gram size).
For the final implementation, we wanted to have a backup metric that works for all input strings.
We filtered edit- and token-based metrics and selected the best candidates according to the criterion described above.
Finally, we selected 27 regular and 4 backup metrics for the second run.
We also added the \textit{equal} metric as a baseline.
We tested those metrics again with $sim^\textit{text} = sim^\textit{code}$, but this time we chose $\vartheta^\textit{\{text, code\}} \in \{0.0, 0.01, 0.02, \ldots, 1.0\}$
Thus, the second run tested 3,232 different configurations, which took about 20 hours.

As motivated above, the results of the text and code metrics depend on each other.
In the third and last run, we tested all combinations of the best (99\% quantile) text and code configurations together with the best backup configurations.
This was the first run with $sim^\textit{text} \ne sim^\textit{code}$ and with a backup metric for text and code blocks.
Those backup metrics were only used if the input strings were too short for the configured metrics.
The run, which took about 14 hours, tested all combinations of 13 text configurations, 3 text backup configurations, 15 code configurations, and 2 code backup configurations, resulting in 1,170 combinations in total.
For the final selection, we ranked the combinations according to the sum of their $MCC$ scores for text and code blocks. The selected configuration was:

\begin{tabbing}
$sim^\textit{text}$ \hspace{0.7em} \= $=~\textit{manhattanFourGramNormalized}$\hspace{2.3em} \= ($\vartheta^\textit{text}\;$ \= $= 0.17$)\\
$sim^\textit{code}$ \> $=~\textit{winnowingFourGramDiceNormalized}$ \> ($\vartheta^\text{code}$ \> $= 0.23$)\\
$sim_\textit{backup}^\textit{text}$ \> $=~\textit{cosineTokenNormalizedTermFrequency}$ \> ($\vartheta^\textit{text}$ \> $= 0.36$)\\
$sim_\textit{backup}^\textit{code}$ \> $=~\textit{cosineTokenNormalizedTermFrequency}$ \> ($\vartheta^\textit{code}$ \> $=  0.26$)
\end{tabbing}

\begin{figure}
\centering
\includegraphics[width=\columnwidth,  trim=0.0in 0.3in 0.2in 0.2in]{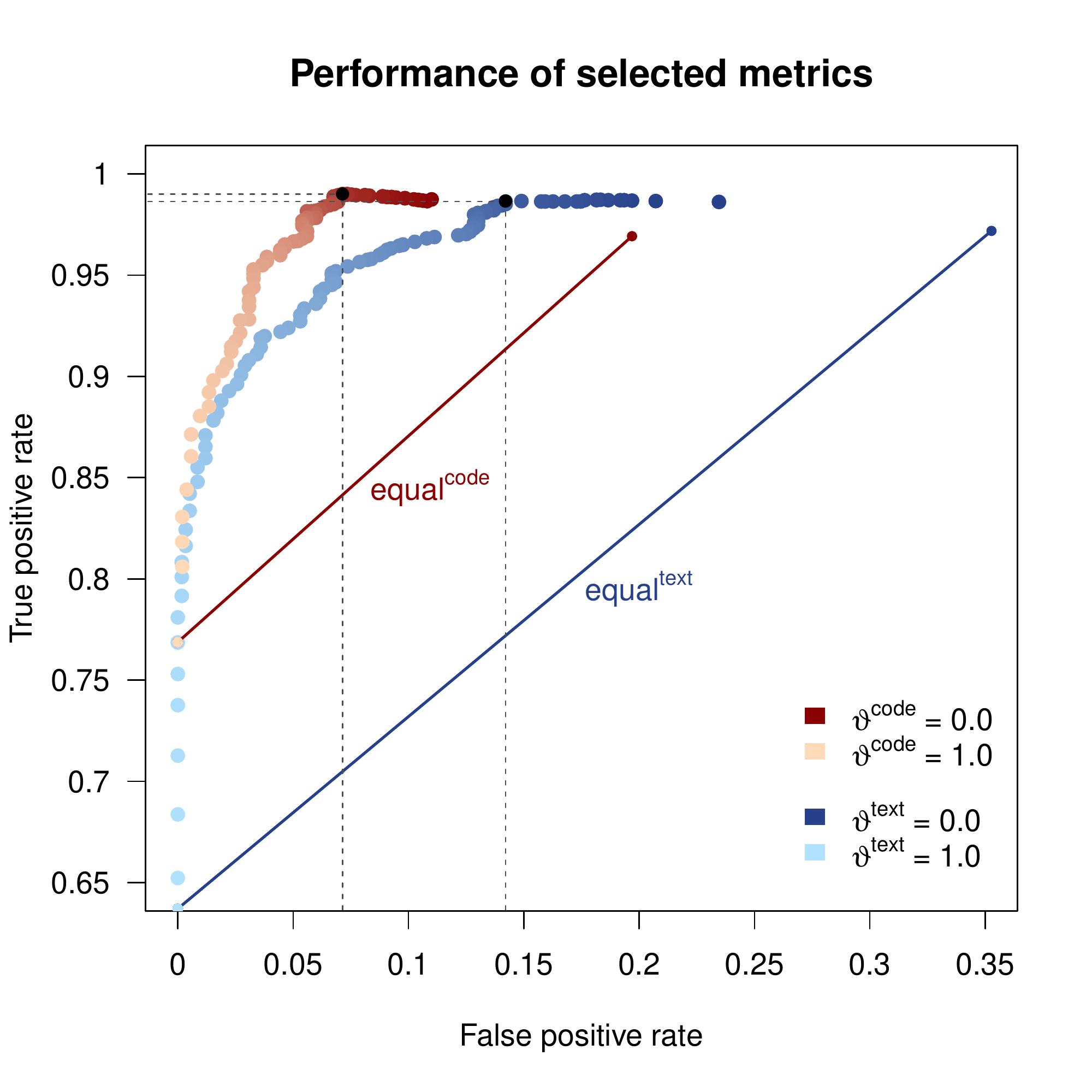} 
\caption{Performance of selected metrics: \textit{manhattanFourGramNormalized} for text (blue) and \textit{winnowingFourGramDiceNormalized} for code (red); selected thresholds: 0.17 for text and 0.23 for code (dotted lines).}
\label{fig:roc-excerpt}
\end{figure}

Figure~\ref{fig:roc-excerpt} shows the performance of the selected metrics for different thresholds with $sim^\textit{text}=sim^\textit{code}$, compared to the baseline metric \textit{equals}.
The final configuration achieved a $MCC$ value of $0.86$ for text (true positive rate $0.99$, false positive rate $0.14$) and $0.92$ for code (true positive rate $0.99$, false positive rate $0.07$).


%
%
%
%
%

\section{Data Analysis}
\label{sec:analysis}


After describing how we reconstructed the version history for individual text and code blocks, we come back to our initial research questions.
We first characterize the phenomenon of SO post evolution, and in particular the evolution of individual post blocks (RQ1).
To find out if edited posts share common characteristics, we analyzed if certain measures such as score or number of comments correlate with the number of edits (RQ2).
We also investigated if those measures have a temporal relationship with the edits, in particular if comments happen immediately before or after edits (RQ3).
As descriptive statistics, we use mean ($M$), standard deviation ($SD$), median ($Mdn$), and the first and third quartiles ($Q_1$, $Q_3$).
To test for significant differences, we applied the nonparametric two-sided \textit{Wilcoxon rank-sum test}~\cite{Wilcoxon1945} and report the corresponding p-value ($p_w$).
To measure the effect size, we used \textit{Cohen's} $d$~\cite{Cohen1988, GibbonsHedekerOthers1993}. 
Our interpretation of $d$ is based on the thresholds described by Cohen~\cite{Cohen1992}: negligible effect ($|d|<0.2$), small effect ($0.2\le|d|<0.5$), medium effect ($0.5\le|d|<0.8$), otherwise large effect.
We used the nonparametric \textit{Spearman's rank correlation coefficient} ($\rho$)~\cite{Spearman1904} to test the statistical dependence between two variables.
Our interpretation of $\rho$ is based on Hinkle et al.'s scheme~\cite{HinkleWiersmaOthers1979}: low correlation ($0.3\leq|\rho|<0.5$), moderate correlation ($0.5\leq|\rho|<0.7$), high correlation ($0.7\leq|\rho|<0.9$), and very high correlation ($0.9\leq|\rho|\leq 1$).

\subsection{Evolution of Stack Overflow Posts}
\label{sec:analysis-evolution}

In the following, we describe different properties of post blocks and post block versions either for their latest version in the dataset, or for different versions over time:

\paragraph{Post Block Count:}
Half of all posts in the \emph{SOTorrent} dataset contain between one and two text blocks and between zero and two code blocks ($Q_{1,3}$). 
There are only few posts without text blocks ($1.0\%$), but over a third of all posts do not have code blocks ($36.6\%$).
Examples for such posts include conceptual questions and answers, but also posts with inline code that we considered to be part of the text blocks.
If we compare the first and the last version of edited posts, we can observe a statistically significant difference in the number of text and code blocks ($p_w^{\textit{text, code}}<\num{2.2e-16}$); posts tend to grow over time.
However, the effect is only small ($d^\textit{text}=0.21,\; d^\textit{code}=0.23$).

\begin{figure}
\centering
\includegraphics[width=0.95\columnwidth,  trim=0.0in 0.8in 0.3in 0.4in]{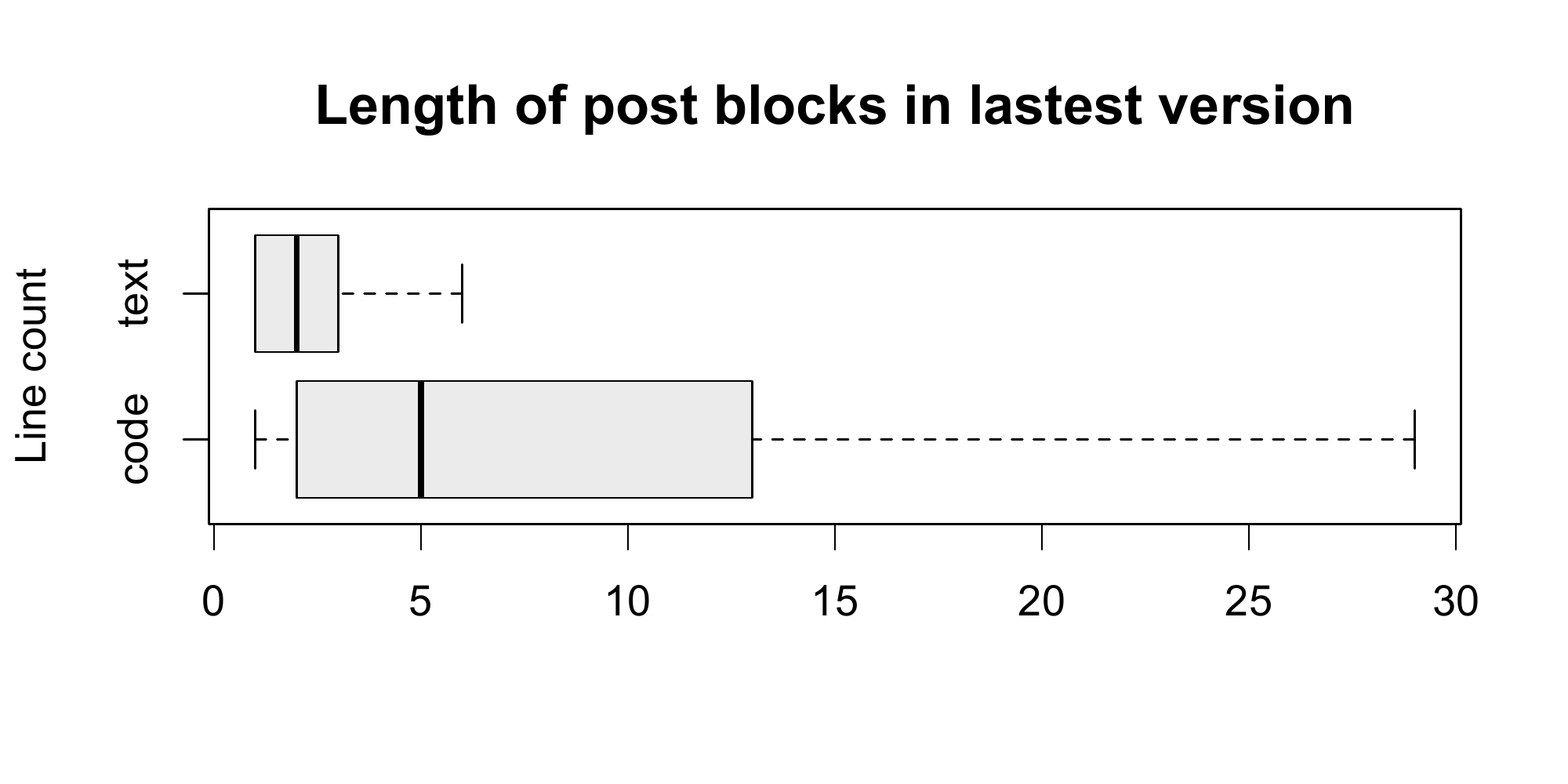} 
\caption{Boxplots showing the line count of text and code blocks in the latest version of Stack Overflow posts ($n=69,940,599$ for text and $n=42,568,011$ for code).}
\label{fig:postblocklength-latest}
\end{figure}

\paragraph{Post Block Length:}
Code blocks tend to be larger than text blocks.
Figure~\ref{fig:postblocklength-latest} visualizes the difference measured in number of lines.
The average text block contains $2.5$ lines ($Mdn=2$, $SD=3.1$) and $247.5$ characters ($Mdn=153$, $SD=319.1$); the average code block contains $12.0$ lines ($Mdn=5$, $SD=23.4$) and $455.9$ characters ($Mdn=194$, $SD=989.3$).
We compared the length of post blocks in the first and the last version and found no effect.
Thus, we can conclude that posts tend to become longer over time in terms of their number of post blocks, but the length of individual post blocks is relatively stable.

\begin{figure}
\centering
\includegraphics[width=0.92\columnwidth,  trim=0.3in 0.9in 0.4in 0.3in]{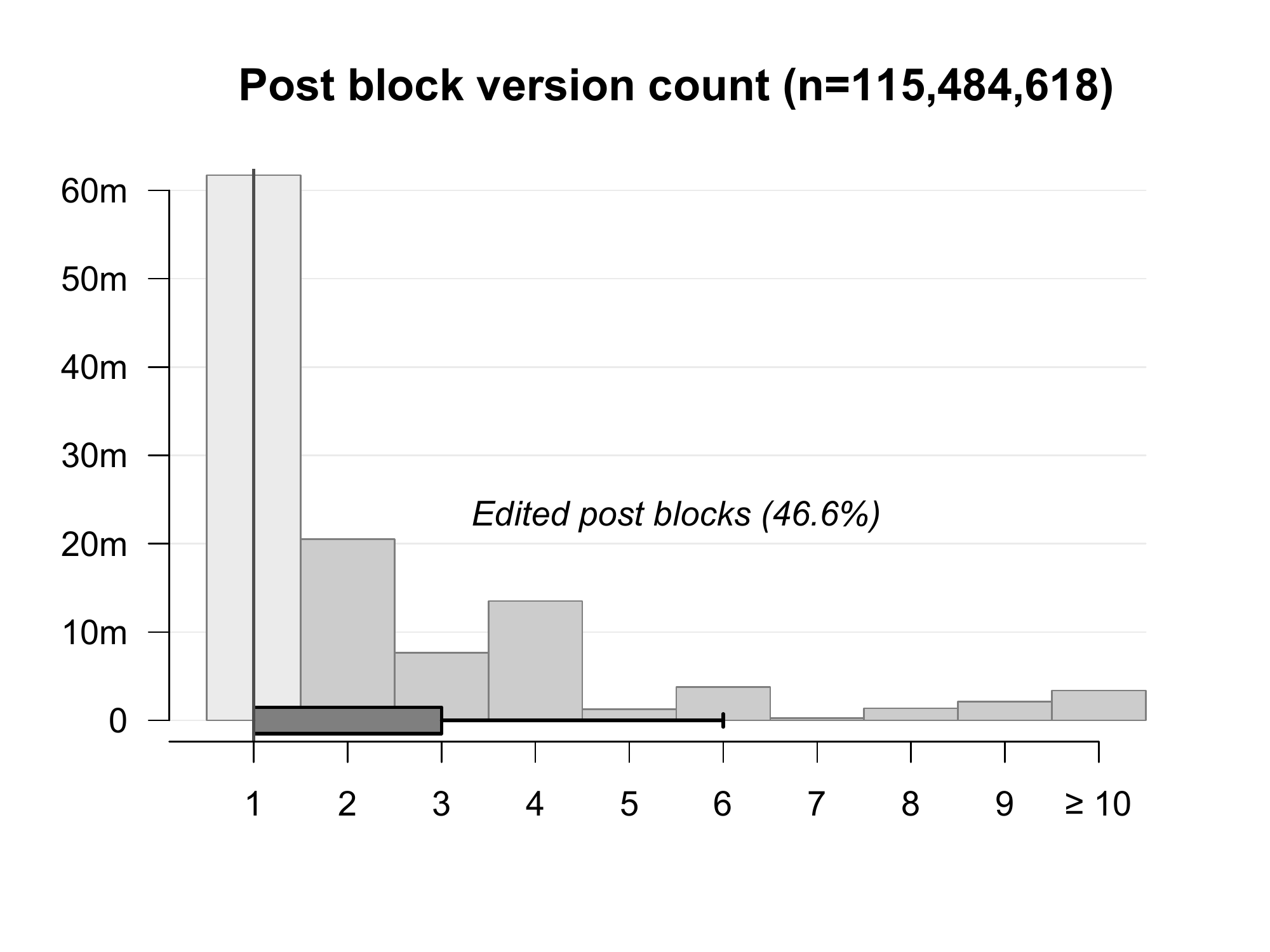} 
\caption{Histogram and boxplot showing the number of post block versions (vertical line is median).}
\label{fig:postblocklifespan-length}
\end{figure}

\paragraph{Post Block Versions:}
For our analysis of post block versions, we retrieved all post block lifespans in the dataset, but only considered the initial versions and later versions where the content of the blocks changed (not all blocks are edited in all versions).
We found that about half of all post blocks were edited after their creation (see Figure~\ref{fig:postblocklifespan-length}).
On average, text blocks have $4.8$ and code blocks $4.1$ versions.
We analyzed the line-based differences between post block versions and found that $44.1\%$ of all edits modify only one line ($47.7\%$ for text blocks and $34.9\%$ for code blocks).
There is a significant difference in the size of changes when comparing text and code blocks ($p_w<\num{2.2e-16}$) with a medium effect ($d=0.51$ for the number of added lines and $d=0.57$ for the number of deleted lines): 
Changes in code blocks are larger, which is expectable due to the larger size of code compared to text blocks.

\paragraph{Post Block Co-change:}
We were also interested in the co-change of text and code blocks, i.e. if text and code is edited together.
On average, $1.5$ text blocks and $0.9$ code blocks were edited in each post version ($Mdn=1$ and $SD=1.1$ for both types).
We found that text and code blocks were either edited together ($49.3\%$ of all post versions), or just the text blocks were edited ($44.6\%$).
Only in $6.1\%$ of all post versions, code blocks were changed without also editing text blocks.
This could indicate that SO authors document changes to their code snippets in the text blocks or update the description of the modified code.

\begin{figure}
\centering
\includegraphics[width=0.95\columnwidth,  trim=0.0in 0.4in 0.3in 0.3in]{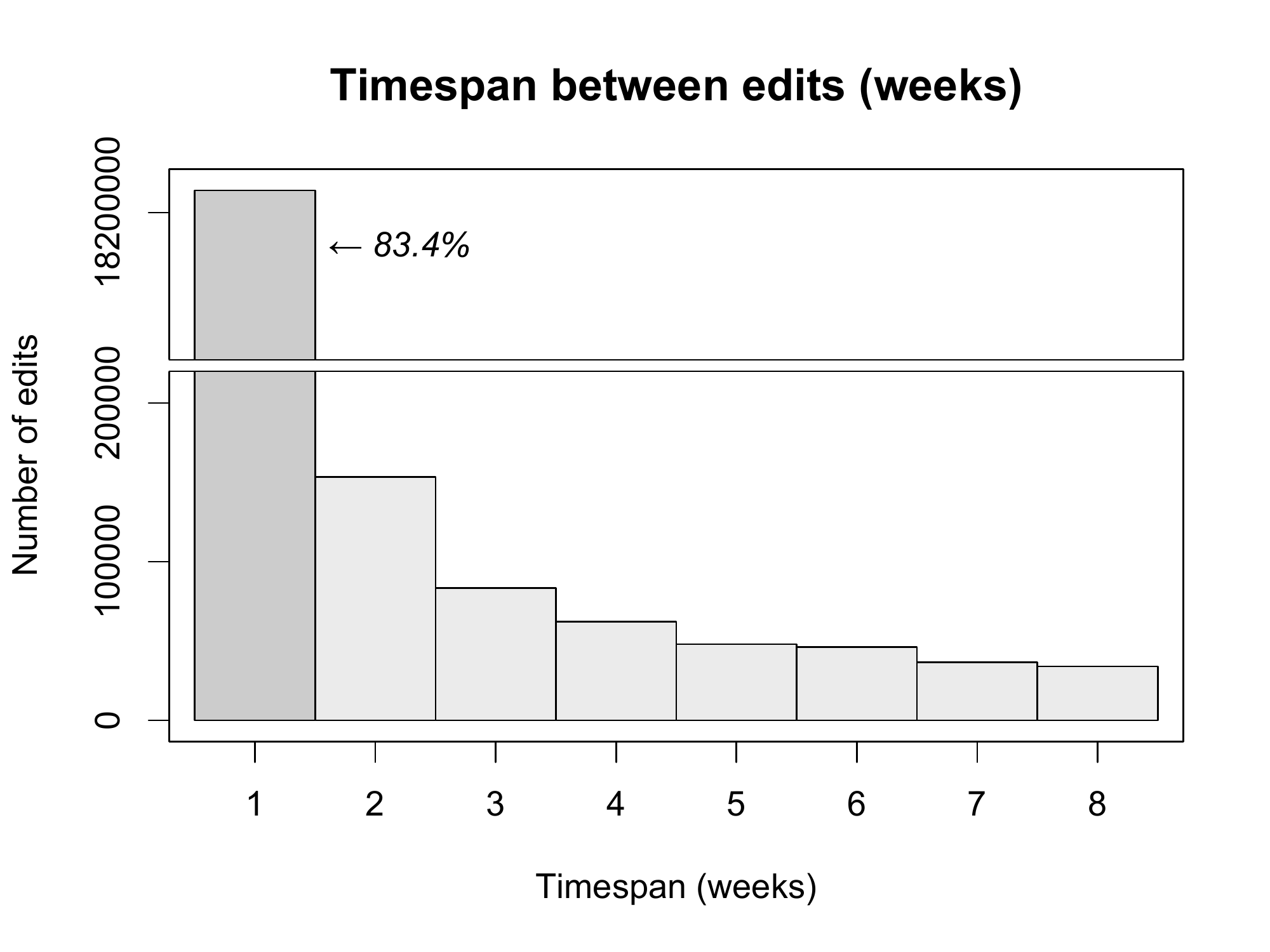} 
\caption{Bar chart visualizing all edit timespans between one and eight weeks ($85.5\%$ of all values, $n=18,677,709$); the other $14.5\%$ are spread over a range of 475 weeks.}
\label{fig:timespan-weeks}
\end{figure}

\paragraph{Order of Post Blocks:}
To check our assumption that the order of post blocks rarely changes, we computed the difference between the local ids of all post blocks versions and their predecessors.
We found that $95.5\%$ of all post block versions have the same local id as their predecessor.
Of all absolute differences, two was the most common one ($3.1\%$), which is expectable, because text and code blocks usually alternate.
Thus, e.g., swapping two blocks of the same type leads to a local id difference of two in the next version.

\paragraph{Timespan Between Edits:}
For the posts that have been edited after their creation, we analyzed the timespan between the edits.
$80.6\%$ of the first post edits happen on the same day as the creation of the post, $4.6\%$ within one week ($>\!1$ and $\le\!7$ days), $5.1\%$ within one year ($>\!7$ and $\le\!365$ days), and $9.7\%$ more than one year after the creation.
If we only consider the second or later edits, not much changes: $74.2\%$ of them happen on the same day, $6.2\%$ within one week, $7.9\%$ within one year, and $11.7\%$ more than one year after the creation.
Overall, $78.2\%$ of all edits happen on the same day, i.e. soon after the creation of the post, and $83.4\%$ happen on the same day or within the first week after the creation (see Figure~\ref{fig:timespan-weeks}).

\paragraph{Post Editors:}
On SO, either the author of a post or a moderator, i.e. a SO user with a reputation of at least 2,000, can make edits.
We found that $87.4\%$ of all edits were conducted by the post authors themselves and $12.6\%$ by moderators.
We found no effect of the authors' reputation on the fact that a moderator edits the post.
We consider an analysis of typical moderator changes to be an interesting direction for future work.

\subsection{Properties of Edited Posts}
\label{sec:properties}

\begin{table}[tb]
\small
\centering
\caption{Correlation table with Spearman's correlation coefficients $\rho~$ for different properties of Stack Overflow posts (p-value $<0.001$ for all combinations).}
\label{tab:correlations}
\vspace{-0.3\baselineskip}
\begin{tabular}{c | rrrrr}
\hline
$\rho$ & \multicolumn{1}{c}{Versions} & \multicolumn{1}{c}{Age} & \multicolumn{1}{c}{Score} & \multicolumn{1}{c}{Comments} & \multicolumn{1}{c}{GHMatches} \\
\hline
Versions &  & $-0.03$ & $0.09$ &\graybg\boldsymbol{$0.26$} & $0.09$ \\
Age & $-0.03$ &  & $0.25$ & $-0.03$ & $0.10$ \\
Score & $0.09$ & $0.25$ &  & $0.08$ & $0.23$ \\
Comments &\graybg\boldsymbol{$0.26$} & $-0.03$ & $0.08$ &  & $0.09$ \\
GHMatches & $0.09$ & $0.10$ & $0.23$ & $0.09$ &  \\
\hline
n & 38.4m & 38.4m & 38.4m & 38.4m & 137k \\
\hline
\end{tabular}
\vspace{-\baselineskip}
\end{table}

To investigate which properties edited posts possess, we searched for monotonic relationships between the version count of a post and other properties such as the age of the post, its score, comment count, or the number of distinct files on GH referring to the post.
Table~\ref{tab:correlations} shows the correlation coefficients ($\rho$) for those relationships.
There was no correlation that exceeded the threshold for a low correlation ($0.3$).
However, the relationship between the version count and the number of comments drew our attention as it had the highest correlation coefficient in the table.
We decided to explore that relationship using a quasi-experiment: We compared the number of comments of all posts with only one version to all posts with more than one version (version count over all posts: $Mdn=1$, $M=1.6$, $SD=1.0$).
The difference was significant ($p_w<\num{2.2e-16}$) and the effect size was medium ($d=0.52$).
We also compared the opposite relationship, i.e. the number of versions of all posts with at most one comment to all posts with more than one comment (comment count over all posts: $Mdn=1$, $M=1.6$, $SD=2.5$).
Again, the difference was significant ($p_w<\num{2.2e-16}$), but the effect size was small ($d=0.49$).

\subsection{Edits and Comments}
\label{sec:edits-comments}

To further explore the relationship between comments and post edits, we looked at their temporal connection, i.e. if comments usually happen before or after edits.
First, we aggregated all edits (including post creation) and all comments per post id and day.
Thus, our units of observation were all days where posts were either created, edited or commented.
We found that in $32.3\%$ of the cases, the posts were created or edited and commented; in $33.3\%$ of the cases they were only created, in $9.1\%$ of the cases only edited, in $7.5\%$ of the cases only created and edited, and in $17.8\%$ of the cases only commented.
If we focus on the comments, we see that $64.4\%$ of them happened on a day where the post had either been created or edited.
We then further focused on those days and calculated the time difference between a comment and the closest edit.
If a comment was closer to the creation then to an edit, we assigned the comment to the creation.
We found that $34.7\%$ of the edits were related to the creation of the post and $65.3\%$ were related to an edit.
Of the latter, $47.9\%$ were made before an edit and $52.1\%$ afterwards. 
Moreover, the comments were usually made right before ($M=-1.2~\text{hours}$, $Mdn=-0.3$, $SD=2.6$) or soon after the edits ($M=+1.3~\text{hours}$, $Mdn=+0.3$, $SD=2.7$).


\section{Discussion}

The \emph{SOTorrent} dataset has allowed us to study the phenomenon of post editing on SO in detail (RQ1).
We found that a total of 13.9 million SO posts ($36.1\%$ of all posts) have been edited at least once.
Many of these edits ($44.1\%$) modify only a single line of text or code, and while posts grow over time in terms of the number of text and code blocks they contain, the size of these individual blocks is relatively stable.
Interestingly, only in $6.1\%$ of all cases are code blocks changed without corresponding changes in text blocks of the same post, suggesting that SO users typically update the textual description accompanying code snippets when they are edited.
Studying the exact nature of such edits will be part of our future work.
We also found that edits are mostly made shortly after the creation of a post ($78.2\%$ of all edits are made on the same day when the post was created), and the vast majority of edits are made by post authors ($87.4\%$)---although the remaining $12.6\%$ will be of particular interest for our future work.
The number of comments on posts without edits is significantly smaller than the number of comments on posts with edits, suggesting an interplay of these two features (RQ2).
We find evidence which suggests that commenting a post on SO helps to bring attention to it (RQ3).
Of the comments that were made on the same day as an edit, $47.9\%$ were made before an edit and $52.1\%$ afterwards, typically (median value) only 18 minutes before or after the edit.
Comments before edits might trigger them, comments after edits might be feedback. 

To investigate the connection between post edits and comments made immediately before or after edits, we conducted a preliminary qualitative analysis.
We drew a random sample of 50 posts, 25 posts for which at least one comment had been made at most 10 minutes before an edit and 25 posts for which at least one comment had been made at most 10 minutes after an edit.
We qualitatively analyzed the posts and found that, in the majority of cases, the comments and edits were clearly related (34 of 50 posts in our sample) and that the edit added or modified a code block (30/50).
We classified a small set of comments as bug reports (10/50) and found that in some cases, the edit was explicitly documented in the post (11/50, e.g., by prefixing content with ``\emph{EDIT:}'').
Comments often asked for additional information (22/50), and in cases where comments happened shortly before the edits, the comment was often a clarifying question (14/25).
Answer 15437937\footnote{https://stackoverflow.com/a/15437937} represents a typical example: In a timespan of 35 minutes, a user answered a question, edited the answer three times, and commented on it once in response to three comments from the user asking the question.
Analyzing such communication structures, e.g., to learn how comments are used for feedback on posts, is part of our future work.

\section{Related Work}

Over the past years, there have been various research papers on leveraging knowledge from SO, e.g., to support post edits~\cite{ChenXingOthers2017}, to automate the search~\citep{PonzanelliBacchelliOthers2013, CampbellTreude2017}, or to augment API documentation~\citep{TreudeRobillard2016}. 
Regarding the population of SO users, studies described properties such as gender~\citep{VasilescuCapiluppiOthers2012} and age~\citep{MorrisonMurphyHill2015}.
Wang et al.~\citep{WangLoDavidOthers2013} analyzed the asking and answering behavior of SO users and found that most of them only answer or ask one question. 
We complement those results with our finding that post edits happen soon after post creation and that comments are closely linked to edits.
Xia et al.~\citep{XiaBaoOthers2017} describe that it is common for developers to search for reusable code snippets on the web.
Yang et al.~\citep{YangHussainOthers2016} found that SO Python and JavaScript snippets are more usable in terms of parsability, compilability and runnability, compared to Java and C\#.
Yang et al.~\citep{YangMartinsOthers2017} analyzed code clones between Python snippets from SO and Python projects on GH and found a considerable number of non-trivial clones, which may have a negative impact on code quality~\citep{AbdalkareemShihabOthers2017}.
Other studies aimed at identifying API usage in SO code snippets~\citep{SubramanianHolmes2015}, describing characteristics of effective code examples~\citep{NasehiSillitoOthers2012}, investigating whether SO code snippets are self-explanatory~\citep{TreudeRobillard2017}, or analyzing the impact of copied SO code snippets on application security~\citep{AcarBackesOthers2016, FischerBottingerOthers2017}.
There has also been work on the interplay between user activity on SO and GH~\citep{VasilescuFilkovOthers2013, SilvestriYangOthers2015, BadashianEstekiOthers2014}.
\emph{SOTorrent} enables researchers to further investigate this connection by collecting links from public GH projects to SO posts. 
To describe topics of SO questions and answers, different methods like manual analysis~\citep{TreudeBarzilayOthers2011} and Latent Dirichlet Allocation~\citep{WangLoDavidOthers2013, AllamanisSutton2015} have been used.
Automatically identifying high-quality posts has been another research direction, where metrics based on the number of edits on a question~\citep{YangHauffOthers2014}, author popularity~\citep{PonzanelliMocciOthers2014}, and code readability~\citep{DuijnKuceraOthers2015} yielded good results.
With our dataset, the evolution of such high-quality posts can easily be analyzed.
German et al.~\citep{GermanDiPentaOthers2009} investigated how code siblings, code clones that evolve in a different system than the original code, flow between systems with different licenses.
Tracing the flow of siblings between GH projects, posts on SO, and external sources is another possible direction for future work that \emph{SOTorrent} can support.
Two fields related to our research are source code plagiarism detection~\citep{LancasterCulwin2004} and code clone detection~\citep{RoyCordyOthers2009}, which both rely on determining the similarity of code fragments. 

\section{Conclusion}

In this paper, we presented \emph{SOTorrent}, an open dataset that enables researchers to analyze the evolution of SO content at the level of whole posts and individual text and code blocks.
We described how we evaluated 134 different string similarity metrics regarding their suitability to match text and code blocks to their predecessor versions.
For text blocks, a profile-based metric using the Manhattan distance yielded the best results; for code blocks, a fingerprint-based metric using the Winnowing algorithm~\cite{SchleimerWilkersonOthers2003, DuricGasevic2013} outperformed the other metrics.
Since multiple predecessor candidates may exist, we also developed a matching strategy that we iteratively refined using random samples of SO posts.
First analyses using the dataset provided new insights into the evolution of SO posts.
In future work, we want to deepen our understanding of how code snippets are maintained on SO.
To this end, we want to identify bug-fixing edits.
Moreover, as \emph{SOTorrent} also collects links from SO posts to other websites and from public GH projects to SO posts, we can explore how code flows from and to external sources like blog posts and open source software projects.
Beside the investigation of new research questions, we will improve and maintain the dataset, for example by developing means to automatically detect code blocks that are not used for code, but for markup (see, e.g., second code block in Figure~\ref{fig:so-postblocks-example}).
Our vision is that \emph{SOTorrent} will help other researchers to further investigate the evolution of SO posts and their connection to other platforms and resources.





\bibliographystyle{ACM-Reference-Format}
\bibliography{literature} 


\begin{thebibliography}{62}


\ifx \showCODEN    \undefined \def \showCODEN     #1{\unskip}     \fi
\ifx \showDOI      \undefined \def \showDOI       #1{#1}\fi
\ifx \showISBNx    \undefined \def \showISBNx     #1{\unskip}     \fi
\ifx \showISBNxiii \undefined \def \showISBNxiii  #1{\unskip}     \fi
\ifx \showISSN     \undefined \def \showISSN      #1{\unskip}     \fi
\ifx \showLCCN     \undefined \def \showLCCN      #1{\unskip}     \fi
\ifx \shownote     \undefined \def \shownote      #1{#1}          \fi
\ifx \showarticletitle \undefined \def \showarticletitle #1{#1}   \fi
\ifx \showURL      \undefined \def \showURL       {\relax}        \fi
\providecommand\bibfield[2]{#2}
\providecommand\bibinfo[2]{#2}
\providecommand\natexlab[1]{#1}
\providecommand\showeprint[2][]{arXiv:#2}

\bibitem[\protect\citeauthoryear{Abdalkareem, Shihab, and Rilling}{Abdalkareem
  et~al\mbox{.}}{2017}]%
        {AbdalkareemShihabOthers2017}
\bibfield{author}{\bibinfo{person}{Rabe Abdalkareem}, \bibinfo{person}{Emad
  Shihab}, {and} \bibinfo{person}{Juergen Rilling}.}
  \bibinfo{year}{2017}\natexlab{}.
\newblock \showarticletitle{{On code reuse from StackOverflow: An exploratory
  study on Android apps}}.
\newblock \bibinfo{journal}{\emph{{Information and Software Technology}}}
  \bibinfo{volume}{88} (\bibinfo{year}{2017}), \bibinfo{pages}{148--158}.
\newblock


\bibitem[\protect\citeauthoryear{Acar, Backes, Fahl, Kim, Mazurek, and
  Stransky}{Acar et~al\mbox{.}}{2016}]%
        {AcarBackesOthers2016}
\bibfield{author}{\bibinfo{person}{Yasemin Acar}, \bibinfo{person}{Michael
  Backes}, \bibinfo{person}{Sascha Fahl}, \bibinfo{person}{Doowon Kim},
  \bibinfo{person}{Michelle~L. Mazurek}, {and} \bibinfo{person}{Christian
  Stransky}.} \bibinfo{year}{2016}\natexlab{}.
\newblock \showarticletitle{{You Get Where You're Looking For: The Impact Of
  Information Sources on Code Security}}. In \bibinfo{booktitle}{\emph{{2016
  IEEE Symposium on Security and Privacy (S{\&}P 2016)}}},
  \bibfield{editor}{\bibinfo{person}{Michael Locasto}, \bibinfo{person}{Vitaly
  Shmatikov}, {and} \bibinfo{person}{{\'U}lfar Erlingsson}} (Eds.).
  \bibinfo{publisher}{{IEEE Computer Society}}, \bibinfo{address}{San Jose, CA,
  USA}, \bibinfo{pages}{289--305}.
\newblock


\bibitem[\protect\citeauthoryear{Allamanis and Sutton}{Allamanis and
  Sutton}{2015}]%
        {AllamanisSutton2015}
\bibfield{author}{\bibinfo{person}{Miltiadis Allamanis} {and}
  \bibinfo{person}{Charles Sutton}.} \bibinfo{year}{2015}\natexlab{}.
\newblock \showarticletitle{{Why, when, and what: Analyzing Stack Overflow
  questions by topic, type, and code}}. In \bibinfo{booktitle}{\emph{{12th
  Working Conference on Mining Software Repositories (MSR 2015)}}},
  \bibfield{editor}{\bibinfo{person}{Massimiliano {Di Penta}},
  \bibinfo{person}{Martin Pinzger}, {and} \bibinfo{person}{Romain Robbes}}
  (Eds.). \bibinfo{publisher}{{IEEE Computer Society}},
  \bibinfo{address}{Florence, Italy}, \bibinfo{pages}{53--56}.
\newblock


\bibitem[\protect\citeauthoryear{An, Mlouki, Khomh, and Antoniol}{An
  et~al\mbox{.}}{2017}]%
        {AnMloukiOthers2017}
\bibfield{author}{\bibinfo{person}{Le.. An}, \bibinfo{person}{Ons Mlouki},
  \bibinfo{person}{Foutse Khomh}, {and} \bibinfo{person}{Giuliano Antoniol}.}
  \bibinfo{year}{2017}\natexlab{}.
\newblock \showarticletitle{{Stack Overflow: A Code Laundering Platform?}}. In
  \bibinfo{booktitle}{\emph{{24th IEEE International Conference on Software
  Analysis, Evolution and Reengineering (SANER 2017)}}},
  \bibfield{editor}{\bibinfo{person}{Martin Pinzger}, \bibinfo{person}{Gabriele
  Bavota}, {and} \bibinfo{person}{Andrian Marcus}} (Eds.).
  \bibinfo{publisher}{{IEEE Computer Society}}, \bibinfo{address}{Klagenfurt,
  Austria}, \bibinfo{pages}{283--293}.
\newblock


\bibitem[\protect\citeauthoryear{Badashian, Esteki, Gholipour, Hindle, and
  Stroulia}{Badashian et~al\mbox{.}}{2014}]%
        {BadashianEstekiOthers2014}
\bibfield{author}{\bibinfo{person}{Ali~Sajedi Badashian},
  \bibinfo{person}{Afsaneh Esteki}, \bibinfo{person}{Ameneh Gholipour},
  \bibinfo{person}{Abram Hindle}, {and} \bibinfo{person}{Eleni Stroulia}.}
  \bibinfo{year}{2014}\natexlab{}.
\newblock \showarticletitle{{Involvement, Contribution and Influence in GitHub
  and Stack Overflow}}. In \bibinfo{booktitle}{\emph{{24th International
  Conference on Computer Science and Software Engineering (CASCON 2014)}}},
  \bibfield{editor}{\bibinfo{person}{Joanna Ng}, \bibinfo{person}{Jin Li},
  {and} \bibinfo{person}{Ken Wong}} (Eds.). \bibinfo{publisher}{{IBM / ACM}},
  \bibinfo{address}{Markham, ON, Canada}, \bibinfo{pages}{19--33}.
\newblock


\bibitem[\protect\citeauthoryear{Baltes}{Baltes}{2018a}]%
        {Baltes2018e}
\bibfield{author}{\bibinfo{person}{Sebastian Baltes}.}
  \bibinfo{year}{2018}\natexlab{a}.
\newblock \bibinfo{title}{{SOTorrent: Reconstructing and Analyzing the
  Evolution of Stack Overflow Posts --- Supplementary Material}}.
\newblock   (\bibinfo{year}{2018}).
\newblock
\urldef\tempurl%
\url{http://doi.org/10.5281/zenodo.1201553}
\showURL{%
\tempurl}


\bibitem[\protect\citeauthoryear{Baltes}{Baltes}{2018b}]%
        {Baltes2018b}
\bibfield{author}{\bibinfo{person}{Sebastian Baltes}.}
  \bibinfo{year}{2018}\natexlab{b}.
\newblock \bibinfo{title}{{sotorrent/db-scripts on GitHub}}.
\newblock   (\bibinfo{year}{2018}).
\newblock
\urldef\tempurl%
\url{https://doi.org/10.5281/zenodo.1116346}
\showURL{%
\tempurl}


\bibitem[\protect\citeauthoryear{Baltes}{Baltes}{2018c}]%
        {Baltes2018d}
\bibfield{author}{\bibinfo{person}{Sebastian Baltes}.}
  \bibinfo{year}{2018}\natexlab{c}.
\newblock \bibinfo{title}{{sotorrent/r-scripts on GitHub}}.
\newblock   (\bibinfo{year}{2018}).
\newblock
\urldef\tempurl%
\url{https://doi.org/10.5281/zenodo.1048185}
\showURL{%
\tempurl}


\bibitem[\protect\citeauthoryear{Baltes and Dumani}{Baltes and Dumani}{2018a}]%
        {BaltesDumani2018}
\bibfield{author}{\bibinfo{person}{Sebastian Baltes} {and}
  \bibinfo{person}{Lorik Dumani}.} \bibinfo{year}{2018}\natexlab{a}.
\newblock \bibinfo{title}{{SOTorrent Data Set Version 2018-02-16}}.
\newblock   (\bibinfo{year}{2018}).
\newblock
\urldef\tempurl%
\url{http://doi.org/10.5281/zenodo.1196296}
\showURL{%
\tempurl}


\bibitem[\protect\citeauthoryear{Baltes and Dumani}{Baltes and Dumani}{2018b}]%
        {BaltesDumani2018b}
\bibfield{author}{\bibinfo{person}{Sebastian Baltes} {and}
  \bibinfo{person}{Lorik Dumani}.} \bibinfo{year}{2018}\natexlab{b}.
\newblock \bibinfo{title}{{sotorrent/metrics-comparison on GitHub}}.
\newblock   (\bibinfo{year}{2018}).
\newblock
\urldef\tempurl%
\url{https://doi.org/10.5281/zenodo.1045823}
\showURL{%
\tempurl}


\bibitem[\protect\citeauthoryear{Baltes and Dumani}{Baltes and Dumani}{2018c}]%
        {BaltesDumani2018c}
\bibfield{author}{\bibinfo{person}{Sebastian Baltes} {and}
  \bibinfo{person}{Lorik Dumani}.} \bibinfo{year}{2018}\natexlab{c}.
\newblock \bibinfo{title}{{sotorrent/so-posthistory-extractor on GitHub}}.
\newblock   (\bibinfo{year}{2018}).
\newblock
\urldef\tempurl%
\url{https://doi.org/10.5281/zenodo.835046}
\showURL{%
\tempurl}


\bibitem[\protect\citeauthoryear{Baltes and Dumani}{Baltes and Dumani}{2018d}]%
        {BaltesDumani2018d}
\bibfield{author}{\bibinfo{person}{Sebastian Baltes} {and}
  \bibinfo{person}{Lorik Dumani}.} \bibinfo{year}{2018}\natexlab{d}.
\newblock \bibinfo{title}{{sotorrent/string-similarity on GitHub}}.
\newblock   (\bibinfo{year}{2018}).
\newblock
\urldef\tempurl%
\url{https://doi.org/10.5281/zenodo.835044}
\showURL{%
\tempurl}


\bibitem[\protect\citeauthoryear{Baltes, Dumani, and Zeimetz}{Baltes
  et~al\mbox{.}}{2017a}]%
        {BaltesDumaniOthers2017}
\bibfield{author}{\bibinfo{person}{Sebastian Baltes}, \bibinfo{person}{Lorik
  Dumani}, {and} \bibinfo{person}{Tobias Zeimetz}.}
  \bibinfo{year}{2017}\natexlab{a}.
\newblock \bibinfo{title}{{Dataset with manually validated version histories of
  Stack Overflow posts}}.
\newblock   (\bibinfo{year}{2017}).
\newblock
\urldef\tempurl%
\url{http://doi.org/10.5281/zenodo.884909}
\showURL{%
\tempurl}


\bibitem[\protect\citeauthoryear{Baltes, Kiefer, and Diehl}{Baltes
  et~al\mbox{.}}{2017b}]%
        {BaltesKieferOthers2017}
\bibfield{author}{\bibinfo{person}{Sebastian Baltes}, \bibinfo{person}{Richard
  Kiefer}, {and} \bibinfo{person}{Stephan Diehl}.}
  \bibinfo{year}{2017}\natexlab{b}.
\newblock \showarticletitle{{Attribution required: Stack overflow code snippets
  in GitHub projects}}. In \bibinfo{booktitle}{\emph{{39th International
  Conference on Software Engineering (ICSE 2017), Companion Volume}}},
  \bibfield{editor}{\bibinfo{person}{Sebasti{\'a}n Uchitel},
  \bibinfo{person}{Alessandro Orso}, {and} \bibinfo{person}{Martin~P.
  Robillard}} (Eds.). \bibinfo{publisher}{{IEEE Computer Society}},
  \bibinfo{address}{Buenos Aires, Argentina}, \bibinfo{pages}{161--163}.
\newblock


\bibitem[\protect\citeauthoryear{Burrows, Tahaghoghi, and Zobel}{Burrows
  et~al\mbox{.}}{2007}]%
        {BurrowsTahaghoghiOthers2007}
\bibfield{author}{\bibinfo{person}{Steven Burrows}, \bibinfo{person}{Seyed
  M.~M. Tahaghoghi}, {and} \bibinfo{person}{Justin Zobel}.}
  \bibinfo{year}{2007}\natexlab{}.
\newblock \showarticletitle{{Efficient plagiarism detection for large code
  repositories}}.
\newblock \bibinfo{journal}{\emph{{Software---Practice and Experience}}}
  \bibinfo{volume}{37}, \bibinfo{number}{2} (\bibinfo{year}{2007}),
  \bibinfo{pages}{151--176}.
\newblock


\bibitem[\protect\citeauthoryear{Campbell and Treude}{Campbell and
  Treude}{2017}]%
        {CampbellTreude2017}
\bibfield{author}{\bibinfo{person}{Brock~Angus Campbell} {and}
  \bibinfo{person}{Christoph Treude}.} \bibinfo{year}{2017}\natexlab{}.
\newblock \showarticletitle{{NLP2Code: Code Snippet Content Assist via Natural
  Language Tasks}}. In \bibinfo{booktitle}{\emph{{2017 IEEE International
  Conference on Software Maintenance and Evolution (ICSME 2017)}}},
  \bibfield{editor}{\bibinfo{person}{Hong Mei}, \bibinfo{person}{Lu~Zhang},
  {and} \bibinfo{person}{Thomas Zimmermann}} (Eds.). \bibinfo{publisher}{{IEEE
  Computer Society}}, \bibinfo{address}{Shanghai, China},
  \bibinfo{pages}{628--632}.
\newblock


\bibitem[\protect\citeauthoryear{Chapin, Hale, Khan, Ramil, and Tan}{Chapin
  et~al\mbox{.}}{2001}]%
        {ChapinHaleOthers2001}
\bibfield{author}{\bibinfo{person}{Ned Chapin}, \bibinfo{person}{Joanne~E.
  Hale}, \bibinfo{person}{Khaled~Md Khan}, \bibinfo{person}{Juan~F. Ramil},
  {and} \bibinfo{person}{Wui-Gee Tan}.} \bibinfo{year}{2001}\natexlab{}.
\newblock \showarticletitle{{Types of software evolution and software
  maintenance}}.
\newblock \bibinfo{journal}{\emph{{Journal of Software Maintenance}}}
  \bibinfo{volume}{13}, \bibinfo{number}{1} (\bibinfo{year}{2001}),
  \bibinfo{pages}{3--30}.
\newblock


\bibitem[\protect\citeauthoryear{Chen, Xing, and Liu}{Chen
  et~al\mbox{.}}{2017}]%
        {ChenXingOthers2017}
\bibfield{author}{\bibinfo{person}{Chunyang Chen}, \bibinfo{person}{Zhenchang
  Xing}, {and} \bibinfo{person}{Yang Liu}.} \bibinfo{year}{2017}\natexlab{}.
\newblock \showarticletitle{{By the Community {\&} For the Community: A Deep
  Learning Approach to Assist Collaborative Editing in Q{\&}A Sites}}.
\newblock \bibinfo{journal}{\emph{{Proceedings of the ACM on Human-Computer
  Interaction}}}  \bibinfo{volume}{1} (\bibinfo{year}{2017}),
  \bibinfo{pages}{32:1--32:21}.
\newblock


\bibitem[\protect\citeauthoryear{Chicco}{Chicco}{2017}]%
        {Chicco2017}
\bibfield{author}{\bibinfo{person}{Davide Chicco}.}
  \bibinfo{year}{2017}\natexlab{}.
\newblock \showarticletitle{{Ten quick tips for machine learning in
  computational biology}}.
\newblock \bibinfo{journal}{\emph{{BioData mining}}} \bibinfo{volume}{10},
  \bibinfo{number}{1} (\bibinfo{year}{2017}), \bibinfo{pages}{35}.
\newblock


\bibitem[\protect\citeauthoryear{Cohen}{Cohen}{1988}]%
        {Cohen1988}
\bibfield{author}{\bibinfo{person}{Jacob Cohen}.}
  \bibinfo{year}{1988}\natexlab{}.
\newblock \bibinfo{booktitle}{\emph{{Statistical Power Analysis for the
  Behavioral Sciences}} (\bibinfo{edition}{2nd} ed.)}.
\newblock \bibinfo{publisher}{Routledge}, \bibinfo{address}{Mahwah, NJ, USA}.
\newblock


\bibitem[\protect\citeauthoryear{Cohen}{Cohen}{1992}]%
        {Cohen1992}
\bibfield{author}{\bibinfo{person}{Jacob Cohen}.}
  \bibinfo{year}{1992}\natexlab{}.
\newblock \showarticletitle{{A power primer}}.
\newblock \bibinfo{journal}{\emph{{Psychological bulletin}}}
  \bibinfo{volume}{112}, \bibinfo{number}{1} (\bibinfo{year}{1992}),
  \bibinfo{pages}{155}.
\newblock


\bibitem[\protect\citeauthoryear{Duijn, Kucera, and Bacchelli}{Duijn
  et~al\mbox{.}}{2015}]%
        {DuijnKuceraOthers2015}
\bibfield{author}{\bibinfo{person}{Maarten Duijn}, \bibinfo{person}{Adam
  Kucera}, {and} \bibinfo{person}{Alberto Bacchelli}.}
  \bibinfo{year}{2015}\natexlab{}.
\newblock \showarticletitle{{Quality Questions Need Quality Code: Classifying
  Code Fragments on Stack Overflow}}. In \bibinfo{booktitle}{\emph{{12th
  Working Conference on Mining Software Repositories (MSR 2015)}}},
  \bibfield{editor}{\bibinfo{person}{Massimiliano {Di Penta}},
  \bibinfo{person}{Martin Pinzger}, {and} \bibinfo{person}{Romain Robbes}}
  (Eds.). \bibinfo{publisher}{{IEEE Computer Society}},
  \bibinfo{address}{Florence, Italy}, \bibinfo{pages}{410--413}.
\newblock


\bibitem[\protect\citeauthoryear{Dumani and Baltes}{Dumani and Baltes}{2017}]%
        {DumaniBaltes2017}
\bibfield{author}{\bibinfo{person}{Lorik Dumani} {and}
  \bibinfo{person}{Sebastian Baltes}.} \bibinfo{year}{2017}\natexlab{}.
\newblock \bibinfo{title}{{sotorrent/so-posthistory-gt on GitHub}}.
\newblock   (\bibinfo{year}{2017}).
\newblock
\urldef\tempurl%
\url{https://doi.org/10.5281/zenodo.1045935}
\showURL{%
\tempurl}


\bibitem[\protect\citeauthoryear{Duric and Gasevic}{Duric and Gasevic}{2013}]%
        {DuricGasevic2013}
\bibfield{author}{\bibinfo{person}{Zoran Duric} {and} \bibinfo{person}{Dragan
  Gasevic}.} \bibinfo{year}{2013}\natexlab{}.
\newblock \showarticletitle{{A source code similarity system for plagiarism
  detection}}.
\newblock \bibinfo{journal}{\emph{{The Computer Journal}}}
  \bibinfo{volume}{56}, \bibinfo{number}{1} (\bibinfo{year}{2013}),
  \bibinfo{pages}{70--86}.
\newblock


\bibitem[\protect\citeauthoryear{Fischer, B{\"o}ttinger, Xiao, Stransky, Acar,
  Backes, and Fahl}{Fischer et~al\mbox{.}}{2017}]%
        {FischerBottingerOthers2017}
\bibfield{author}{\bibinfo{person}{Felix Fischer}, \bibinfo{person}{Konstantin
  B{\"o}ttinger}, \bibinfo{person}{Huang Xiao}, \bibinfo{person}{Christian
  Stransky}, \bibinfo{person}{Yasemin Acar}, \bibinfo{person}{Michael Backes},
  {and} \bibinfo{person}{Sascha Fahl}.} \bibinfo{year}{2017}\natexlab{}.
\newblock \showarticletitle{{Stack Overflow Considered Harmful? The Impact of
  Copy{\&}Paste on Android Application Security}}. In
  \bibinfo{booktitle}{\emph{{2017 IEEE Symposium on Security and Privacy
  (S{\&}P 2017)}}}, \bibfield{editor}{\bibinfo{person}{Kevin R.~B. Butler},
  \bibinfo{person}{{\'U}lfar Erlingsson}, {and} \bibinfo{person}{Bryan Parno}}
  (Eds.). \bibinfo{publisher}{{IEEE Computer Society}}, \bibinfo{address}{San
  Jose, CA, USA}, \bibinfo{pages}{121--136}.
\newblock


\bibitem[\protect\citeauthoryear{German, {Di Penta}, Gueheneuc, and
  Antoniol}{German et~al\mbox{.}}{2009}]%
        {GermanDiPentaOthers2009}
\bibfield{author}{\bibinfo{person}{Daniel~M. German},
  \bibinfo{person}{Massimiliano {Di Penta}}, \bibinfo{person}{Yann-Gael
  Gueheneuc}, {and} \bibinfo{person}{Giuliano Antoniol}.}
  \bibinfo{year}{2009}\natexlab{}.
\newblock \showarticletitle{{Code siblings: Technical and legal implications of
  copying code between applications}}. In \bibinfo{booktitle}{\emph{{6th
  International Working Conference on Mining Software Repositories (MSR
  2009)}}}, \bibfield{editor}{\bibinfo{person}{Michael~W. Godfrey} {and}
  \bibinfo{person}{Jim Whitehead}} (Eds.). \bibinfo{publisher}{{IEEE Computer
  Society}}, \bibinfo{address}{Vancouver, BC, Canada}, \bibinfo{pages}{81--90}.
\newblock


\bibitem[\protect\citeauthoryear{Gharehyazie, Ray, and Filkov}{Gharehyazie
  et~al\mbox{.}}{2017}]%
        {GharehyazieRayOthers2017}
\bibfield{author}{\bibinfo{person}{Mohammad Gharehyazie},
  \bibinfo{person}{Baishakhi Ray}, {and} \bibinfo{person}{Vladimir Filkov}.}
  \bibinfo{year}{2017}\natexlab{}.
\newblock \showarticletitle{{Some From Here, Some From There: Cross-Project
  Code Reuse in GitHub}}. In \bibinfo{booktitle}{\emph{{14th International
  Conference on Mining Software Repositories (MSR 2017)}}},
  \bibfield{editor}{\bibinfo{person}{Jesus~M. Gonzalez-Barahona},
  \bibinfo{person}{Abram Hindle}, {and} \bibinfo{person}{Lin Tan}} (Eds.).
  \bibinfo{publisher}{{IEEE Computer Society}}, \bibinfo{address}{Buenos Aires,
  Argentina}, \bibinfo{pages}{291--301}.
\newblock


\bibitem[\protect\citeauthoryear{Gibbons, Hedeker, and Davis}{Gibbons
  et~al\mbox{.}}{1993}]%
        {GibbonsHedekerOthers1993}
\bibfield{author}{\bibinfo{person}{Robert~D. Gibbons},
  \bibinfo{person}{Donald~R. Hedeker}, {and} \bibinfo{person}{John~M. Davis}.}
  \bibinfo{year}{1993}\natexlab{}.
\newblock \showarticletitle{{Estimation of effect size from a series of
  experiments involving paired comparisons}}.
\newblock \bibinfo{journal}{\emph{{Journal of Educational Statistics}}}
  \bibinfo{volume}{18}, \bibinfo{number}{3} (\bibinfo{year}{1993}),
  \bibinfo{pages}{271--279}.
\newblock


\bibitem[\protect\citeauthoryear{Godfrey and German}{Godfrey and
  German}{2008}]%
        {GodfreyGerman2008}
\bibfield{author}{\bibinfo{person}{Michael~W. Godfrey} {and}
  \bibinfo{person}{Daniel~M. German}.} \bibinfo{year}{2008}\natexlab{}.
\newblock \showarticletitle{{The past, present, and future of software
  evolution}}. In \bibinfo{booktitle}{\emph{{Frontiers of Software Maintenance
  (FoSM 2008)}}}, \bibfield{editor}{\bibinfo{person}{Hausi Muller},
  \bibinfo{person}{Scott Tilley}, {and} \bibinfo{person}{Kenny Wong}} (Eds.).
  \bibinfo{publisher}{IEEE}, \bibinfo{address}{Beijing, China},
  \bibinfo{pages}{129--138}.
\newblock


\bibitem[\protect\citeauthoryear{{Google Cloud Platform}}{{Google Cloud
  Platform}}{2018}]%
        {GoogleCloudPlatform2018}
\bibfield{author}{\bibinfo{person}{{Google Cloud Platform}}.}
  \bibinfo{year}{2018}\natexlab{}.
\newblock \bibinfo{title}{{GitHub Data}}.
\newblock   (\bibinfo{year}{2018}).
\newblock
\urldef\tempurl%
\url{https://cloud.google.com/bigquery/public-data/github}
\showURL{%
\tempurl}


\bibitem[\protect\citeauthoryear{Gousios}{Gousios}{2013}]%
        {Gousios2013}
\bibfield{author}{\bibinfo{person}{Georgios Gousios}.}
  \bibinfo{year}{2013}\natexlab{}.
\newblock \showarticletitle{{The GHTorrent dataset and tool suite}}. In
  \bibinfo{booktitle}{\emph{{10th International Working Conference on Mining
  Software Repositories (MSR 2013)}}},
  \bibfield{editor}{\bibinfo{person}{Thomas Zimmermann},
  \bibinfo{person}{Massimiliano {Di Penta}}, {and} \bibinfo{person}{Sunghun
  Kim}} (Eds.). \bibinfo{publisher}{IEEE}, \bibinfo{address}{San Francisco, CA,
  USA}, \bibinfo{pages}{233--236}.
\newblock


\bibitem[\protect\citeauthoryear{Hinkle, Wiersma, and Jurs}{Hinkle
  et~al\mbox{.}}{1979}]%
        {HinkleWiersmaOthers1979}
\bibfield{author}{\bibinfo{person}{Dennis~E. Hinkle}, \bibinfo{person}{William
  Wiersma}, {and} \bibinfo{person}{Stephen~G. Jurs}.}
  \bibinfo{year}{1979}\natexlab{}.
\newblock \bibinfo{booktitle}{\emph{{Applied statistics for the behavioral
  sciences}}}.
\newblock \bibinfo{publisher}{{Rand McNally College Publishing}},
  \bibinfo{address}{Skokie, IL, USA}.
\newblock


\bibitem[\protect\citeauthoryear{Lancaster and Culwin}{Lancaster and
  Culwin}{2004}]%
        {LancasterCulwin2004}
\bibfield{author}{\bibinfo{person}{Thomas Lancaster} {and}
  \bibinfo{person}{Fintan Culwin}.} \bibinfo{year}{2004}\natexlab{}.
\newblock \showarticletitle{{A comparison of source code plagiarism detection
  engines}}.
\newblock \bibinfo{journal}{\emph{{Computer Science Education}}}
  \bibinfo{volume}{14}, \bibinfo{number}{2} (\bibinfo{year}{2004}),
  \bibinfo{pages}{101--112}.
\newblock


\bibitem[\protect\citeauthoryear{Lehman}{Lehman}{1980}]%
        {Lehman1980}
\bibfield{author}{\bibinfo{person}{Meir~M. Lehman}.}
  \bibinfo{year}{1980}\natexlab{}.
\newblock \showarticletitle{{Programs, life cycles, and laws of software
  evolution}}.
\newblock \bibinfo{journal}{\emph{{Proceedings of the IEEE}}}
  \bibinfo{volume}{68}, \bibinfo{number}{9} (\bibinfo{year}{1980}),
  \bibinfo{pages}{1060--1076}.
\newblock


\bibitem[\protect\citeauthoryear{Manning, Raghavan, and Schutze}{Manning
  et~al\mbox{.}}{2008}]%
        {ManningRaghavanOthers2008}
\bibfield{author}{\bibinfo{person}{Christopher~D. Manning},
  \bibinfo{person}{Prabhakar Raghavan}, {and} \bibinfo{person}{Hinrich
  Schutze}.} \bibinfo{year}{2008}\natexlab{}.
\newblock \bibinfo{booktitle}{\emph{{Introduction to Information Retrieval}}}.
\newblock \bibinfo{publisher}{{Cambridge University Press}},
  \bibinfo{address}{New York, NY, USA}.
\newblock


\bibitem[\protect\citeauthoryear{Martins, Fonte, Henriques, and Cruz}{Martins
  et~al\mbox{.}}{2014}]%
        {MartinsFonteOthers2014}
\bibfield{author}{\bibinfo{person}{Vitor~T. Martins}, \bibinfo{person}{Daniela
  Fonte}, \bibinfo{person}{Pedro~Rangel Henriques}, {and}
  \bibinfo{person}{Daniela~da Cruz}.} \bibinfo{year}{2014}\natexlab{}.
\newblock \showarticletitle{{Plagiarism Detection: A Tool Survey and
  Comparison}}. In \bibinfo{booktitle}{\emph{{3rd Symposium on Languages,
  Applications and Technologies (SLATE 2014)}}}
  \emph{(\bibinfo{series}{OpenAccess Series in Informatics (OASIcs)})},
  \bibfield{editor}{\bibinfo{person}{Maria Jo{\~a}o~Varanda Pereira},
  \bibinfo{person}{Jos{\'e}~Paulo Leal}, {and} \bibinfo{person}{Alberto
  Simoes}} (Eds.), Vol.~\bibinfo{volume}{38}. \bibinfo{publisher}{{Schloss
  Dagstuhl--Leibniz-Zentrum fuer Informatik}},
  \bibinfo{address}{Bragan{\c{c}}a, Portugal}, \bibinfo{pages}{143--158}.
\newblock


\bibitem[\protect\citeauthoryear{Matthews}{Matthews}{1975}]%
        {Matthews1975}
\bibfield{author}{\bibinfo{person}{Brian~W. Matthews}.}
  \bibinfo{year}{1975}\natexlab{}.
\newblock \showarticletitle{{Comparison of the predicted and observed secondary
  structure of T4 phage lysozyme}}.
\newblock \bibinfo{journal}{\emph{{Biochimica et Biophysica Acta (BBA) --
  Protein Structure}}} \bibinfo{volume}{405}, \bibinfo{number}{2}
  (\bibinfo{year}{1975}), \bibinfo{pages}{442--451}.
\newblock


\bibitem[\protect\citeauthoryear{Mens and Demeyer}{Mens and Demeyer}{2008}]%
        {MensDemeyer2008}
\bibfield{editor}{\bibinfo{person}{Tom Mens} {and} \bibinfo{person}{Serge
  Demeyer}} (Eds.). \bibinfo{year}{2008}\natexlab{}.
\newblock \bibinfo{booktitle}{\emph{{Software Evolution}}}.
\newblock \bibinfo{publisher}{Springer}, \bibinfo{address}{Berlin, Germany}.
\newblock


\bibitem[\protect\citeauthoryear{Morrison and Murphy-Hill}{Morrison and
  Murphy-Hill}{2015}]%
        {MorrisonMurphyHill2015}
\bibfield{author}{\bibinfo{person}{Patrick Morrison} {and}
  \bibinfo{person}{Emerson Murphy-Hill}.} \bibinfo{year}{2015}\natexlab{}.
\newblock \showarticletitle{{Is programming knowledge related to age? An
  exploration of Stack Overflow}}. In \bibinfo{booktitle}{\emph{{12th Working
  Conference on Mining Software Repositories (MSR 2015)}}},
  \bibfield{editor}{\bibinfo{person}{Massimiliano {Di Penta}},
  \bibinfo{person}{Martin Pinzger}, {and} \bibinfo{person}{Romain Robbes}}
  (Eds.). \bibinfo{publisher}{{IEEE Computer Society}},
  \bibinfo{address}{Florence, Italy}, \bibinfo{pages}{69--72}.
\newblock


\bibitem[\protect\citeauthoryear{Nasehi, Sillito, Maurer, and Burns}{Nasehi
  et~al\mbox{.}}{2012}]%
        {NasehiSillitoOthers2012}
\bibfield{author}{\bibinfo{person}{Seyed~Mehdi Nasehi},
  \bibinfo{person}{Jonathan Sillito}, \bibinfo{person}{Frank Maurer}, {and}
  \bibinfo{person}{Chris Burns}.} \bibinfo{year}{2012}\natexlab{}.
\newblock \showarticletitle{{What makes a good code example? A study of
  programming Q{\&}A in StackOverflow}}. In \bibinfo{booktitle}{\emph{{28th
  IEEE International Conference on Software Maintenance (ICSM 2012)}}},
  \bibfield{editor}{\bibinfo{person}{Paolo Tonella},
  \bibinfo{person}{Massimiliano {Di Penta}}, {and} \bibinfo{person}{Jonathan~I.
  Maletic}} (Eds.). \bibinfo{publisher}{{IEEE Computer Society}},
  \bibinfo{address}{Trento, Italy}, \bibinfo{pages}{25--34}.
\newblock


\bibitem[\protect\citeauthoryear{Ponzanelli, Bacchelli, and Lanza}{Ponzanelli
  et~al\mbox{.}}{2013}]%
        {PonzanelliBacchelliOthers2013}
\bibfield{author}{\bibinfo{person}{Luca Ponzanelli}, \bibinfo{person}{Alberto
  Bacchelli}, {and} \bibinfo{person}{Michele Lanza}.}
  \bibinfo{year}{2013}\natexlab{}.
\newblock \showarticletitle{{Seahawk: Stack Overflow in the IDE}}. In
  \bibinfo{booktitle}{\emph{{35th International Conference on Software
  Engineering (ICSE 2013)}}}, \bibfield{editor}{\bibinfo{person}{David Notkin},
  \bibinfo{person}{Betty H.~C. Cheng}, {and} \bibinfo{person}{Klaus Pohl}}
  (Eds.). \bibinfo{publisher}{{IEEE Computer Society}}, \bibinfo{address}{San
  Francisco, CA, USA}, \bibinfo{pages}{1295--1298}.
\newblock


\bibitem[\protect\citeauthoryear{Ponzanelli, Mocci, Bacchelli, and
  Lanza}{Ponzanelli et~al\mbox{.}}{2014}]%
        {PonzanelliMocciOthers2014}
\bibfield{author}{\bibinfo{person}{Luca Ponzanelli}, \bibinfo{person}{Andrea
  Mocci}, \bibinfo{person}{Alberto Bacchelli}, {and} \bibinfo{person}{Michele
  Lanza}.} \bibinfo{year}{2014}\natexlab{}.
\newblock \showarticletitle{{Understanding and classifying the quality of
  technical forum questions}}. In \bibinfo{booktitle}{\emph{{14th International
  Conference on Quality Software (QSIC 2014)}}},
  \bibfield{editor}{\bibinfo{person}{W.~Eric Wong} {and} \bibinfo{person}{Bruce
  McMillin}} (Eds.). \bibinfo{publisher}{IEEE}, \bibinfo{address}{Allen, TX,
  USA}, \bibinfo{pages}{343--352}.
\newblock


\bibitem[\protect\citeauthoryear{Powers}{Powers}{2011}]%
        {Powers2011}
\bibfield{author}{\bibinfo{person}{David~Martin Powers}.}
  \bibinfo{year}{2011}\natexlab{}.
\newblock \showarticletitle{{Evaluation: From precision, recall and F-measure
  to ROC, informedness, markedness and correlation}}.
\newblock \bibinfo{journal}{\emph{{Journal of Machine Learning Technologies}}}
  \bibinfo{volume}{2}, \bibinfo{number}{1} (\bibinfo{year}{2011}),
  \bibinfo{pages}{37--63}.
\newblock


\bibitem[\protect\citeauthoryear{Roy, Cordy, and Koschke}{Roy
  et~al\mbox{.}}{2009}]%
        {RoyCordyOthers2009}
\bibfield{author}{\bibinfo{person}{Chanchal~K. Roy}, \bibinfo{person}{James~R.
  Cordy}, {and} \bibinfo{person}{Rainer Koschke}.}
  \bibinfo{year}{2009}\natexlab{}.
\newblock \showarticletitle{{Comparison and evaluation of code clone detection
  techniques and tools: A qualitative approach}}.
\newblock \bibinfo{journal}{\emph{{Science of Computer Programming}}}
  \bibinfo{volume}{74}, \bibinfo{number}{7} (\bibinfo{year}{2009}),
  \bibinfo{pages}{470--495}.
\newblock


\bibitem[\protect\citeauthoryear{Schleimer, Wilkerson, and Aiken}{Schleimer
  et~al\mbox{.}}{2003}]%
        {SchleimerWilkersonOthers2003}
\bibfield{author}{\bibinfo{person}{Saul Schleimer}, \bibinfo{person}{Daniel~S.
  Wilkerson}, {and} \bibinfo{person}{Alex Aiken}.}
  \bibinfo{year}{2003}\natexlab{}.
\newblock \showarticletitle{{Winnowing: Local algorithms for document
  fingerprinting}}. In \bibinfo{booktitle}{\emph{{2003 ACM SIGMOD International
  Conference on Management of Data (SIGMOD 2003)}}},
  \bibfield{editor}{\bibinfo{person}{Alon~Y. Halevy},
  \bibinfo{person}{Zachary~G. Ives}, {and} \bibinfo{person}{AnHai Doan}}
  (Eds.). \bibinfo{publisher}{ACM}, \bibinfo{address}{San Diego, CA, USA},
  \bibinfo{pages}{76--85}.
\newblock


\bibitem[\protect\citeauthoryear{Silvestri, Yang, Bozzon, and
  Tagarelli}{Silvestri et~al\mbox{.}}{2015}]%
        {SilvestriYangOthers2015}
\bibfield{author}{\bibinfo{person}{Giuseppe Silvestri}, \bibinfo{person}{Jie
  Yang}, \bibinfo{person}{Alessandro Bozzon}, {and} \bibinfo{person}{Andrea
  Tagarelli}.} \bibinfo{year}{2015}\natexlab{}.
\newblock \showarticletitle{{Linking Accounts across Social Networks: The Case
  of StackOverflow, GitHub and Twitter}}. In \bibinfo{booktitle}{\emph{{1st
  International Workshop on Knowledge Discovery on the WEB (KDWeb 2015)}}}
  \emph{(\bibinfo{series}{CEUR Workshop Proceedings})},
  \bibfield{editor}{\bibinfo{person}{Giuliano Armano},
  \bibinfo{person}{Alessandro Bozzon}, {and} \bibinfo{person}{Alessandro
  Giuliani}} (Eds.). \bibinfo{publisher}{CEUR-WS.org},
  \bibinfo{address}{Cagliari, Italy}, \bibinfo{pages}{41--52}.
\newblock


\bibitem[\protect\citeauthoryear{Spearman}{Spearman}{1904}]%
        {Spearman1904}
\bibfield{author}{\bibinfo{person}{Charles Spearman}.}
  \bibinfo{year}{1904}\natexlab{}.
\newblock \showarticletitle{{The proof and measurement of association between
  two things}}.
\newblock \bibinfo{journal}{\emph{{American Journal of Psychology}}}
  \bibinfo{volume}{15}, \bibinfo{number}{1} (\bibinfo{year}{1904}),
  \bibinfo{pages}{72--101}.
\newblock


\bibitem[\protect\citeauthoryear{{Stack Exchange Community Wiki}}{{Stack
  Exchange Community Wiki}}{2 27}]%
        {StackExchangeCommunityWiki2018}
\bibfield{author}{\bibinfo{person}{{Stack Exchange Community Wiki}}.}
  \bibinfo{year}{2018-02-27}\natexlab{}.
\newblock \bibinfo{title}{{Database schema documentation for the public data
  dump and SEDE}}.
\newblock   (\bibinfo{year}{2018-02-27}).
\newblock
\urldef\tempurl%
\url{https://meta.stackexchange.com/a/2678}
\showURL{%
\tempurl}


\bibitem[\protect\citeauthoryear{{Stack Exchange Inc}}{{Stack Exchange
  Inc}}{2017}]%
        {StackExchangeInc2017b}
\bibfield{author}{\bibinfo{person}{{Stack Exchange Inc}}.}
  \bibinfo{year}{2017}\natexlab{}.
\newblock \bibinfo{title}{{Stack Exchange Data Dump 2017-12-01}}.
\newblock   (\bibinfo{year}{2017}).
\newblock
\urldef\tempurl%
\url{https://archive.org/details/stackexchange/}
\showURL{%
\tempurl}


\bibitem[\protect\citeauthoryear{{Stack Exchange Inc}}{{Stack Exchange
  Inc}}{2018}]%
        {StackExchangeInc2018}
\bibfield{author}{\bibinfo{person}{{Stack Exchange Inc}}.}
  \bibinfo{year}{2018}\natexlab{}.
\newblock \bibinfo{title}{{Markdown help}}.
\newblock   (\bibinfo{year}{2018}).
\newblock
\urldef\tempurl%
\url{https://stackoverflow.com/editing-help}
\showURL{%
\tempurl}


\bibitem[\protect\citeauthoryear{Subramanian and Holmes}{Subramanian and
  Holmes}{2015}]%
        {SubramanianHolmes2015}
\bibfield{author}{\bibinfo{person}{Siddharth Subramanian} {and}
  \bibinfo{person}{Reid Holmes}.} \bibinfo{year}{2015}\natexlab{}.
\newblock \showarticletitle{{Making sense of online code snippets}}. In
  \bibinfo{booktitle}{\emph{{12th Working Conference on Mining Software
  Repositories (MSR 2015)}}}, \bibfield{editor}{\bibinfo{person}{Massimiliano
  {Di Penta}}, \bibinfo{person}{Martin Pinzger}, {and} \bibinfo{person}{Romain
  Robbes}} (Eds.). \bibinfo{publisher}{{IEEE Computer Society}},
  \bibinfo{address}{Florence, Italy}, \bibinfo{pages}{85--88}.
\newblock


\bibitem[\protect\citeauthoryear{Treude, Barzilay, and Storey}{Treude
  et~al\mbox{.}}{2011}]%
        {TreudeBarzilayOthers2011}
\bibfield{author}{\bibinfo{person}{Christoph Treude}, \bibinfo{person}{Ohad
  Barzilay}, {and} \bibinfo{person}{Margaret-Anne~D. Storey}.}
  \bibinfo{year}{2011}\natexlab{}.
\newblock \showarticletitle{{How do programmers ask and answer questions on the
  web?}}. In \bibinfo{booktitle}{\emph{{33rd International Conference on
  Software Engineering (ICSE 2011)}}},
  \bibfield{editor}{\bibinfo{person}{Richard~N. Taylor},
  \bibinfo{person}{Harald~C. Gall}, {and} \bibinfo{person}{Nenad Medvidovic}}
  (Eds.). \bibinfo{publisher}{ACM}, \bibinfo{address}{Waikiki, Honolulu},
  \bibinfo{pages}{804--807}.
\newblock


\bibitem[\protect\citeauthoryear{Treude and Robillard}{Treude and
  Robillard}{2016}]%
        {TreudeRobillard2016}
\bibfield{author}{\bibinfo{person}{Christoph Treude} {and}
  \bibinfo{person}{Martin~P. Robillard}.} \bibinfo{year}{2016}\natexlab{}.
\newblock \showarticletitle{{Augmenting API Documentation with Insights from
  Stack Overflow}}. In \bibinfo{booktitle}{\emph{{38th International Conference
  on Software Engineering (ICSE 2016)}}},
  \bibfield{editor}{\bibinfo{person}{Laura Dillon}, \bibinfo{person}{Willem
  Visser}, {and} \bibinfo{person}{Laurie Williams}} (Eds.).
  \bibinfo{publisher}{ACM}, \bibinfo{address}{Austin, TX, USA},
  \bibinfo{pages}{392--403}.
\newblock


\bibitem[\protect\citeauthoryear{Treude and Robillard}{Treude and
  Robillard}{2017}]%
        {TreudeRobillard2017}
\bibfield{author}{\bibinfo{person}{Christoph Treude} {and}
  \bibinfo{person}{Martin~P. Robillard}.} \bibinfo{year}{2017}\natexlab{}.
\newblock \showarticletitle{{Understanding Stack Overflow Code Fragments}}. In
  \bibinfo{booktitle}{\emph{{2017 IEEE International Conference on Software
  Maintenance and Evolution (ICSME 2017)}}},
  \bibfield{editor}{\bibinfo{person}{Hong Mei}, \bibinfo{person}{Lu~Zhang},
  {and} \bibinfo{person}{Thomas Zimmermann}} (Eds.). \bibinfo{publisher}{{IEEE
  Computer Society}}, \bibinfo{address}{Shanghai, China},
  \bibinfo{pages}{509--513}.
\newblock


\bibitem[\protect\citeauthoryear{Vasilescu, Capiluppi, and
  Serebrenik}{Vasilescu et~al\mbox{.}}{2012}]%
        {VasilescuCapiluppiOthers2012}
\bibfield{author}{\bibinfo{person}{Bogdan Vasilescu}, \bibinfo{person}{Andrea
  Capiluppi}, {and} \bibinfo{person}{Alexander Serebrenik}.}
  \bibinfo{year}{2012}\natexlab{}.
\newblock \showarticletitle{{Gender, Representation and Online Participation: A
  Quantitative Study of StackOverflow}}. In \bibinfo{booktitle}{\emph{{4th
  International Conference on Social Informatics (SocInfo 2012)}}}
  \emph{(\bibinfo{series}{Lecture Notes in Computer Science})},
  \bibfield{editor}{\bibinfo{person}{Karl Aberer}, \bibinfo{person}{Andreas
  Flache}, \bibinfo{person}{Wander Jager}, \bibinfo{person}{Ling Liu},
  \bibinfo{person}{Jie Tang}, {and} \bibinfo{person}{Christophe Gueret}}
  (Eds.). \bibinfo{publisher}{Springer}, \bibinfo{address}{Lausanne,
  Switzerland}, \bibinfo{pages}{332--338}.
\newblock


\bibitem[\protect\citeauthoryear{Vasilescu, Filkov, and Serebrenik}{Vasilescu
  et~al\mbox{.}}{2013}]%
        {VasilescuFilkovOthers2013}
\bibfield{author}{\bibinfo{person}{Bogdan Vasilescu}, \bibinfo{person}{Vladimir
  Filkov}, {and} \bibinfo{person}{Alexander Serebrenik}.}
  \bibinfo{year}{2013}\natexlab{}.
\newblock \showarticletitle{{StackOverflow and GitHub: Associations between
  Software Development and Crowdsourced Knowledge}}. In
  \bibinfo{booktitle}{\emph{{2013 International Conference on Social Computing
  (SocialCom 2013)}}}, \bibfield{editor}{\bibinfo{person}{L.~W. Chang},
  \bibinfo{person}{Jaideep Srivastava}, {and} \bibinfo{person}{Justin Zhan}}
  (Eds.). \bibinfo{publisher}{{IEEE Computer Society}},
  \bibinfo{address}{Washington, DC, USA}, \bibinfo{pages}{188--195}.
\newblock


\bibitem[\protect\citeauthoryear{Wang, {Lo David}, and Jiang}{Wang
  et~al\mbox{.}}{2013}]%
        {WangLoDavidOthers2013}
\bibfield{author}{\bibinfo{person}{Shaowei Wang}, \bibinfo{person}{{Lo David}},
  {and} \bibinfo{person}{Lingxiao Jiang}.} \bibinfo{year}{2013}\natexlab{}.
\newblock \showarticletitle{{An empirical study on developer interactions in
  StackOverflow}}. In \bibinfo{booktitle}{\emph{{28th Annual ACM Symposium on
  Applied Computing (SAC 2013)}}}, \bibfield{editor}{\bibinfo{person}{Sung~Y.
  Shin} {and} \bibinfo{person}{Jos{\'e}~Carlos Maldonado}} (Eds.).
  \bibinfo{publisher}{ACM}, \bibinfo{address}{Coimbra, Portugal},
  \bibinfo{pages}{1019--1024}.
\newblock


\bibitem[\protect\citeauthoryear{Wilcoxon}{Wilcoxon}{1945}]%
        {Wilcoxon1945}
\bibfield{author}{\bibinfo{person}{Frank Wilcoxon}.}
  \bibinfo{year}{1945}\natexlab{}.
\newblock \showarticletitle{{Individual comparisons by ranking methods}}.
\newblock \bibinfo{journal}{\emph{{Biometrics}}} \bibinfo{volume}{1},
  \bibinfo{number}{6} (\bibinfo{year}{1945}), \bibinfo{pages}{80--83}.
\newblock


\bibitem[\protect\citeauthoryear{Xia, Bao, Lo, Kochhar, Hassan, and Xing}{Xia
  et~al\mbox{.}}{2017}]%
        {XiaBaoOthers2017}
\bibfield{author}{\bibinfo{person}{Xin Xia}, \bibinfo{person}{Lingfeng Bao},
  \bibinfo{person}{David Lo}, \bibinfo{person}{Pavneet~Singh Kochhar},
  \bibinfo{person}{Ahmed~E. Hassan}, {and} \bibinfo{person}{Zhenchang Xing}.}
  \bibinfo{year}{2017}\natexlab{}.
\newblock \showarticletitle{{What do developers search for on the web?}}
\newblock \bibinfo{journal}{\emph{{Empirical Software Engineering}}}
  \bibinfo{volume}{22}, \bibinfo{number}{6} (\bibinfo{year}{2017}),
  \bibinfo{pages}{3149--3185}.
\newblock


\bibitem[\protect\citeauthoryear{Yang, Hussain, and Lopes}{Yang
  et~al\mbox{.}}{2016}]%
        {YangHussainOthers2016}
\bibfield{author}{\bibinfo{person}{Di. Yang}, \bibinfo{person}{Aftab Hussain},
  {and} \bibinfo{person}{Cristina~Videira Lopes}.}
  \bibinfo{year}{2016}\natexlab{}.
\newblock \showarticletitle{{From Query to Usable Code: An Analysis of Stack
  Overflow Code Snippets}}. In \bibinfo{booktitle}{\emph{{13th International
  Conference on Mining Software Repositories (MSR 2016)}}},
  \bibfield{editor}{\bibinfo{person}{Miryung Kim}, \bibinfo{person}{Romain
  Robbes}, {and} \bibinfo{person}{Christian Bird}} (Eds.).
  \bibinfo{publisher}{ACM}, \bibinfo{address}{Austin, TX, USA},
  \bibinfo{pages}{391--402}.
\newblock


\bibitem[\protect\citeauthoryear{Yang, Martins, Saini, and Lopes}{Yang
  et~al\mbox{.}}{2017}]%
        {YangMartinsOthers2017}
\bibfield{author}{\bibinfo{person}{Di. Yang}, \bibinfo{person}{Pedro Martins},
  \bibinfo{person}{Vaibhav Saini}, {and} \bibinfo{person}{Cristina~V. Lopes}.}
  \bibinfo{year}{2017}\natexlab{}.
\newblock \showarticletitle{{Stack Overflow in Github: Any Snippets There?}}.
  In \bibinfo{booktitle}{\emph{{14th International Conference on Mining
  Software Repositories (MSR 2017)}}},
  \bibfield{editor}{\bibinfo{person}{Jesus~M. Gonzalez-Barahona},
  \bibinfo{person}{Abram Hindle}, {and} \bibinfo{person}{Lin Tan}} (Eds.).
  \bibinfo{publisher}{{IEEE Computer Society}}, \bibinfo{address}{Buenos Aires,
  Argentina}, \bibinfo{pages}{280--290}.
\newblock


\bibitem[\protect\citeauthoryear{Yang, Hauff, Bozzon, and Houben}{Yang
  et~al\mbox{.}}{2014}]%
        {YangHauffOthers2014}
\bibfield{author}{\bibinfo{person}{Jie Yang}, \bibinfo{person}{Claudia Hauff},
  \bibinfo{person}{Alessandro Bozzon}, {and} \bibinfo{person}{Geert-Jan
  Houben}.} \bibinfo{year}{2014}\natexlab{}.
\newblock \showarticletitle{{Asking the right question in collaborative Q{\&}A
  systems}}. In \bibinfo{booktitle}{\emph{{25th ACM Conference on Hypertext and
  Social Media (HT 2014)}}}, \bibfield{editor}{\bibinfo{person}{Leo Ferres},
  \bibinfo{person}{Gustavo Rossi}, \bibinfo{person}{Virgilio A.~F. Almeida},
  {and} \bibinfo{person}{Eelco Herder}} (Eds.). \bibinfo{publisher}{ACM},
  \bibinfo{address}{Santiago, Chile}, \bibinfo{pages}{179--189}.
\newblock
\showISBNx{978-1-4503-2954-5}


\end{thebibliography}

\end{document}